\documentclass[prd,aps,epsfig,amsmath]{revtex4}

\usepackage{graphicx}
\usepackage{dcolumn}
\usepackage{bm}

\begin{document}

\title{Renormalization of the baryon axial vector current in large-$N_c$ chiral perturbation theory
}

\author{
Rub\'en Flores-Mendieta
}
\affiliation{
Instituto de F{\'\i}sica, Universidad Aut\'onoma de San Luis Potos{\'\i}, \'Alvaro Obreg\'on 64, Zona Centro, San Luis Potos{\'\i}, S.L.P.\ 78000, M\'exico
}

\author{Christoph P.\ Hofmann
}
\affiliation{
Facultad de Ciencias, Universidad de Colima, Bernal D{\'\i}az del Castillo 340, Colima, Colima 28045, M\'exico
}

\date{\today}

\begin{abstract}
The baryon axial vector current is computed at one-loop order in heavy baryon chiral perturbation theory in the large-$N_c$ limit, where $N_c$ is the number of colors. Loop
graphs with octet and decuplet intermediate states cancel to various orders in $N_c$ as a consequence of the large-$N_c$ spin-flavor symmetry of QCD baryons. These
cancellations are explicitly shown for the general case of $N_f$ flavors of light quarks. In particular, a new generic cancellation is identified in the renormalization of
the baryon axial vector current at one-loop order. A comparison with conventional heavy baryon chiral perturbation theory is performed at the physical values
$N_c\!=\!3, N_f\!=\!3$.
\end{abstract}

\pacs{12.39.Fe, 11.15.Pg,12.38.Bx}

\maketitle

\section{Introduction}
\label{intro}

Despite the tremendous progress achieved in the understanding of the strong interactions with Quantum Chromodynamics, analytic calculations of the spectrum
and properties of hadrons are not possible because the theory is strongly coupled at low energies, with no small expansion parameter. One thus has to resort to the
implementation of alternative methods in order to extract low-energy consequences of QCD. Among these methods, chiral perturbation theory and the $1/N_c$ expansion
(where $N_c$ is the number of colors) have shed much light on the subject.

On the one hand, chiral perturbation theory exploits the symmetry of the QCD Lagrangian under $SU(3)_L\times SU(3)_R\times U(1)_V$ transformations of the three flavors of
light quarks in the limit $m_q\to 0$. Chiral symmetry is spontaneously broken by the QCD vacuum to the vector subgroup $SU(3)_V\times U(1)_V$, giving rise to an octet of
Goldstone bosons. Physical observables can be expanded order by order in powers of $p^2/{\Lambda_\chi}^2$ and $m_{\Pi}^2/{\Lambda_\chi}^2$, or equivalently,
$m_q/{\Lambda_\chi}$, where $p$ is the meson momentum, $m_{\Pi}$ is the mass of the Goldstone boson and $\Lambda_\chi$ is the scale of chiral symmetry breaking. When chiral
perturbation theory is extended to include baryons, it is convenient to introduce velocity-dependent baryon fields, so that the expansion of the baryon chiral Lagrangian in
powers of $m_q$ and $1/M_B$ (where $M_B$ is the baryon mass) is manifest \cite{jm255,jm259}. This so-called heavy baryon chiral perturbation theory was first
applied to compute the chiral logarithmic corrections to the baryon axial vector current for baryon semileptonic decays due to meson loops \cite{jm255,jm259}.
While these corrections are large when only octet baryon intermediate states are kept \cite{jm255}, the inclusion of decuplet baryon intermediate states yields
sizable cancellations between one-loop corrections \cite{jm259}. This phenomenological observation can be rigorously explained in the context of the $1/N_c$
expansion \cite{dm315,j315,djm94} and will be illustrated in detail in the present paper for the case of the baryon axial vector current.

On the other hand, the generalization of QCD from $N_c\!=\!3$ to $N_c \gg 3$ colors, known as the large-$N_c$ limit, has also led to remarkable insights into the
understanding of the nonperturbative QCD dynamics of hadrons. In the large-$N_c$ limit the meson sector of QCD consists of a spectrum of narrow resonances and
meson-meson scattering amplitudes are suppressed by powers of $1/\sqrt{N_c}$ \cite{th}. The baryon sector of QCD, on the contrary, is more subtle to analyze
\cite{witten} because in the large-$N_c$ limit an exact contracted $SU(2N_f)$ spin-flavor symmetry (where $N_f$ is the number of light quark flavors) emerges
\cite{dm315,gs}. This symmetry can be used to classify large-$N_c$ baryon states and matrix elements. It is then possible to consider physical quantities in the
large-$N_c$ limit, where corrections arise at relative orders $1/N_c$, $1/N_c^2$ and so on, which is precisely the origin of the $1/N_c$ expansion. Applications of
this formalism to the computation of static properties of baryons range from masses \cite{djm94,djm95,jl}, couplings \cite{djm94,djm95,dai,rfm98} to magnetic
moments \cite{dai,lb}, to name but a few.

In the present paper, we use a combined expansion in $m_q$ and $1/N_c$. The $1/N_c$ chiral effective Lagrangian for the lowest-lying baryons was constructed in
Refs.~\cite{jen96,lmr94} and describes the interactions of the spin-$\frac{1}{2}$ baryon octet and the spin-$\frac{3}{2}$ baryon decuplet with the pion nonet. Within
this framework we then compute the renormalization of the baryon axial vector current at the one-loop level. As already pointed out in Refs.~\cite{dm315,j315,djm94,fmhjm},
there are large-$N_c$ cancellations between individual Feynman diagrams, provided one sums over all baryon states in a complete multiplet of the large-$N_c$ $SU(6)$
spin-flavor symmetry, i.e., over both the octet and decuplet, and uses axial coupling ratios given by the large-$N_c$ spin-flavor symmetry. In Ref.~\cite{fmhjm}
the general structure of the various large-$N_c$ cancellations was analyzed. In particular, a new large-$N_c$ cancellation was identified. Our work goes beyond
this global analysis as we explicitly evaluate the corresponding operator expressions that involve complicated structures of commutators and/or anticommutators of
$SU(6)$ spin-flavor operators. Although straightforward in principle, the reduction of these operator products to a physical operator basis turns out to be quite tedious
due to the considerable amount of group theory involved. Our final expressions explicitly demonstrate how these large-$N_c$ cancellations occur. In
particular, we show that the new large-$N_c$ cancellation found in Ref.~\cite{fmhjm} is a generic feature of the corresponding commutator-anticommutator
structure and not just occurs in the special case considered in this reference.

Our analysis also contains a comparison of the results obtained within the framework of large-$N_c$ baryon chiral perturbation theory with conventional heavy baryon chiral
perturbation theory (including both octet and decuplet baryons), where no $1/N_c$ expansion is involved. Both approaches agree -- the large-$N_c$ cancellations are
guaranteed to occur as a consequence of the contracted $SU(6)$ spin-flavor symmetry present in the limit $N_c \to \infty$: No large numerical cancellations between loop
diagrams with intermediate octet states and low-energy constants of the next-to-leading order effective Lagrangian, containing the effects of decuplet states, arise.

The present paper is organized as follows. In Sec.~\ref{large-N chiral} we give a brief overview of the structure of the $1/N_c$ chiral effective Lagrangian for the
lowest-lying baryons. In order to make the paper self-contained, Sec.~\ref{large-N chiral} also contains the relevant large-$N_c$ formalism. The renormalization of the
baryon axial vector current is considered in Sec.~\ref{renormalization}. Here, we present in detail our basic calculation, i.e., the reduction of complicated structures of
commutators and/or anticommutators, and show how the various large-$N_c$ cancellations occur. Formulas for the physically interesting case of three colors and three light
quark flavors are given explicitly. In Sec.~\ref{HBCHPT} we discuss the renormalization of the baryon axial vector current within the framework of heavy baryon chiral
perturbation theory in a form that allows us to then make the comparison with large-$N_c$ baryon chiral perturbation theory in Sec.~\ref{comparison}; we close this latter
section by performing a fit to the experimental data on baryon semileptonic decays. The inclusion of the $\eta^\prime$, which becomes a Goldstone boson in the limit
$N_c \to \infty$, is performed in Sec.~\ref{etaPrime}. Finally, we present our conclusions in Sec.~\ref{conclusions}. The paper contains three appendices. In Appendix A
the most general expressions for the complicated reduced structures of commutators and/or anticommutators are given for an arbitrary number of colors and light quark flavors.
Appendix B contains tables of matrix elements of spin-flavor operators relevant to discuss eight observed transitions between spin-$\frac{1}{2}$ baryons. In particular, we
illustrate how one extracts the axial vector couplings for the semileptonic processes of physical interest. Finally, Appendix C lists the chiral coefficients occurring in
the renormalization of the baryon axial vector current.

\section{The chiral Lagrangian for baryons in the $1/N_c$ expansion}
\label{large-N chiral}

The formalism of heavy baryon chiral perturbation theory and the $1/N_c$ baryon chiral Lagrangian have been discussed in detail in Ref.~\cite{jen96}. In this section we
restrict ourselves to presenting an overview and introducing our notation and conventions.

The $1/N_c$ baryon chiral Lagrangian which correctly implements nonet symmetry and contracted spin-flavor symmetry for baryons in the large-$N_c$ limit can be
written in the most general way as
\begin{equation}
\mathcal{L}_{\text{baryon}} = i \mathcal{D}^0 - \mathcal{M}_{\text{hyperfine}} + \text{Tr} \left(\mathcal{A}^k \lambda^c \right) A^{kc} + \frac{1}{N_c} \text{Tr}
\left(\mathcal{A}^k \frac{2I}{\sqrt 6}\right) A^k + \ldots, \label{eq:ncch}
\end{equation}
where
\begin{equation}
\mathcal{D}^0 = \partial^0 \openone + \text{Tr} \left(\mathcal{V}^0 \lambda^c\right) T^c. \label{eq:kin}
\end{equation}
Each term in Eq.~(\ref{eq:ncch}) involves a baryon operator which can be expressed as a polynomial in the $SU(6)$ spin-flavor generators \cite{djm95}
\begin{equation}
J^k = q^\dagger \frac{\sigma^k}{2} q, \qquad T^c = q^\dagger \frac{\lambda^c}{2} q, \qquad G^{kc} = q^\dagger
\frac{\sigma^k}{2}\frac{\lambda^c}{2} q, \label{eq:su6gen}
\end{equation}
where $q^\dagger$ and $q$ are $SU(6)$ operators that create and annihilate states in the fundamental representation of $SU(6)$, and
$\sigma^k$ and $\lambda^c$ are the Pauli spin and Gell-Mann flavor matrices, respectively. In Eqs.(\ref{eq:ncch})-(\ref{eq:su6gen}) the flavor indices
run from one to nine so the full meson nonet $\pi$, $K$, $\eta$, and $\eta^\prime$ is considered.

The baryon operator $\mathcal{M}_{\text{hyperfine}}$ denotes the spin splittings of the tower of baryon states with spins $1/2,\ldots, N_c/2$ in the flavor
representations. Furthermore, the vector and axial vector combinations of the meson fields,
\begin{equation}
\mathcal{V}^0 = \frac12 \left(\xi \partial^0 \xi^\dagger + \xi^\dagger \partial^0 \xi\right), \qquad \qquad
\mathcal{A}^k = \frac{i}{2} \left(\xi \nabla^k \xi^\dagger - \xi^\dagger \nabla^k \xi\right),
\end{equation}
couple to baryon vector and axial vector currents, respectively. Here $\xi=\exp[i\Pi(x)/f]$, where $\Pi(x)$ stands for the nonet of Goldstone boson fields (unless explicitly
stated otherwise) and $f \approx 93$ MeV is the meson decay constant. In particular, the $\ell = 1$ flavor octet axial vector pion combination couples to the flavor octet
baryon axial vector current, denoted by $A^{kc}$ hereafter.

The QCD operators involved in $\mathcal{L}_{\text{baryon}}$ in Eq.~(\ref{eq:ncch}) have well-defined $1/N_c$ expansions. Specifically, the baryon axial vector
current $A^{kc}$ is a spin-1 object, an octet under $SU(3)$, and odd under time reversal. Its $1/N_c$ expansion can be written as \cite{djm95}
\begin{equation}
A^{kc} = a_1 G^{kc} + \sum_{n=2,3}^{N_c} b_n \frac{1}{N_c^{n-1}} \mathcal{D}_n^{kc} + \sum_{n=3,5}^{N_c} c_n
\frac{1}{N_c^{n-1}} \mathcal{O}_n^{kc}, \label{eq:akcfull}
\end{equation}
where the $\mathcal{D}_n^{kc}$ are diagonal operators with nonzero matrix elements only between states with the same spin, and the $\mathcal{O}_n^{kc}$ are purely
off-diagonal operators with nonzero matrix elements only between states with different spin. The first few terms in expansion (\ref{eq:akcfull}) read
\begin{eqnarray}
\mathcal{D}_2^{kc} & = & J^kT^c, \label{eq:d2kc} \\
\mathcal{O}_2^{kc} & = & \epsilon^{ijk} \{J^i,G^{jc}\}, \label{eq:o2kc} \\
\mathcal{D}_3^{kc} & = & \{J^k,\{J^r,G^{rc}\}\}, \label{eq:d3kc} \\
\mathcal{O}_3^{kc} & = & \{J^2,G^{kc}\} - \frac12 \{J^k,\{J^r,G^{rc}\}\}. \label{eq:o3kc}
\end{eqnarray}
Higher order terms can be obtained via $\mathcal{D}_n^{kc}=\{J^2,\mathcal{D}_{n-2}^{kc}\}$ and $\mathcal{O}_n^{kc}=\{J^2,\mathcal{O}_{n-2}^{kc}\}$ for $n\geq 4$.
From the above definitions it is easy to verify that the operators $\mathcal{O}_{2m}^{kc}$ $(m=1,2,\ldots)$ are forbidden in the expansion (\ref{eq:akcfull})
because they are even under time reversal. Furthermore, the unknown coefficients $a_1$, $b_n$, and $c_n$ in Eq.~(\ref{eq:akcfull}) have expansions in powers of
$1/N_c$ and are order unity at leading order in the $1/N_c$ expansion. At the physical value $N_c = 3$ the series can be truncated as
\begin{equation}
A^{kc} = a_1 G^{kc} + b_2 \frac{1}{N_c} \mathcal{D}_2^{kc} + b_3 \frac{1}{N_c^2} \mathcal{D}_3^{kc} + c_3 \frac{1}{N_c^2} \mathcal{O}_3^{kc}. \label{eq:akc}
\end{equation}
The matrix elements of the space components of $A^{kc}$ between $SU(6)$ symmetric states give the actual values of the axial vector couplings. For the octet baryons,
the axial vector couplings are $g_A$, as conventionally defined in baryon $\beta$-decay experiments, with a normalization such that $g_A \approx 1.27$ and $g_V=1$
for neutron decay.

Similarly, the baryon axial current $A^k$ is a spin-1 object, a singlet under $SU(3)$ so its $1/N_c$ expansion can be written as \cite{jen96}
\begin{equation}
A^k = \sum_{n=1,3}^{N_c} b_n^{1,1} \frac{1}{N_c^{n-1}} \mathcal{D}_n^k, \label{eq:asin}
\end{equation}
where $\mathcal{D}_1^k = J^k$ and $\mathcal{D}_{2m+1}^k = \{J^2,\mathcal{D}_{2m-1}^k\}$ for $m\geq 1$. The superscript on the operator coefficients of $A^k$ denotes that
they refer to the baryon singlet current. For $N_c=3$, Eq.~(\ref{eq:asin}) reduces to
\begin{equation}
A^k = b_1^{1,1} J^k + b_3^{1,1} \frac{1}{N_c^2} \{J^2,J^k\}.
\end{equation}

As for the baryon mass operator $\mathcal{M}$, its $1/N_c$ expansion can be written as \cite{dm315,djm94,djm95,lmr94}
\begin{eqnarray}
\mathcal{M} = m_0 N_c \openone + \sum_{n=2,4}^{N_c-1} m_{n} \frac{1}{N_c^{n-1}} J^n, \label{eq:mop}
\end{eqnarray}
where $m_n$ are unknown coefficients. The first term on the right-hand side of Eq.~(\ref{eq:mop}) is the overall spin-independent mass
of the baryon multiplet
and is removed from the chiral Lagrangian by the heavy baryon field redefinition~\cite{jm255}. The remaining terms are spin-dependent and define
$\mathcal{M}_{\text{hyperfine}}$ introduced in Eq.~(\ref{eq:ncch}). For $N_c=3$ the hyperfine mass expansion reduces to a single operator
\begin{eqnarray}
\mathcal{M} _{\text{hyperfine}} = \frac{m_2}{N_c} J^2 . \label{eq:smop}
\end{eqnarray}

\section{Renormalization of the baryon axial vector current}
\label{renormalization}

One of the earliest applications of Lagrangian (\ref{eq:ncch}) consisted in the calculation of nonanalytic meson-loop corrections in Ref.~\cite{jen96}. Specifically,
the calculation of the flavor \textbf{27} contribution to the baryon masses was presented in this reference as an example.

The renormalization of the baryon axial vector current is another problem which can be analyzed within the formalism of Ref.~\cite{jen96}. Aspects of this problem
have been discussed in the framework of heavy baryon chiral perturbation theory \cite{jm255,jm259,b2}, the $1/N_c$ expansion \cite{djm94,djm95,dai} or in a
simultaneous expansion in chiral symmetry breaking and $1/N_c$ \cite{weise,fmhjm}. This latter approach is implemented in the present work to the calculation of the
renormalization of the baryon axial vector current at one-loop order, following the lines of Ref.~\cite{jen96}. There are, however, some aspects of this problem which
have not been previously discussed and will be addressed here.

The baryon axial vector current $A^{kc}$ is renormalized by the one-loop diagrams displayed in Fig.~\ref{fig:eins}. These loop graphs have a calculable dependence
on the ratio $\Delta/m_\Pi$, where $\Delta \equiv M_\Delta - M_N$ is the decuplet-octet mass difference and $m_\Pi$ is the meson mass. Let us discuss the diagrams
of Figs.~\ref{fig:eins}(a,b,c) and Fig.~\ref{fig:eins}(d) separately, as they involve different commutator-anticommutator structures.

\begin{figure}[ht]
\scalebox{0.9}{\includegraphics{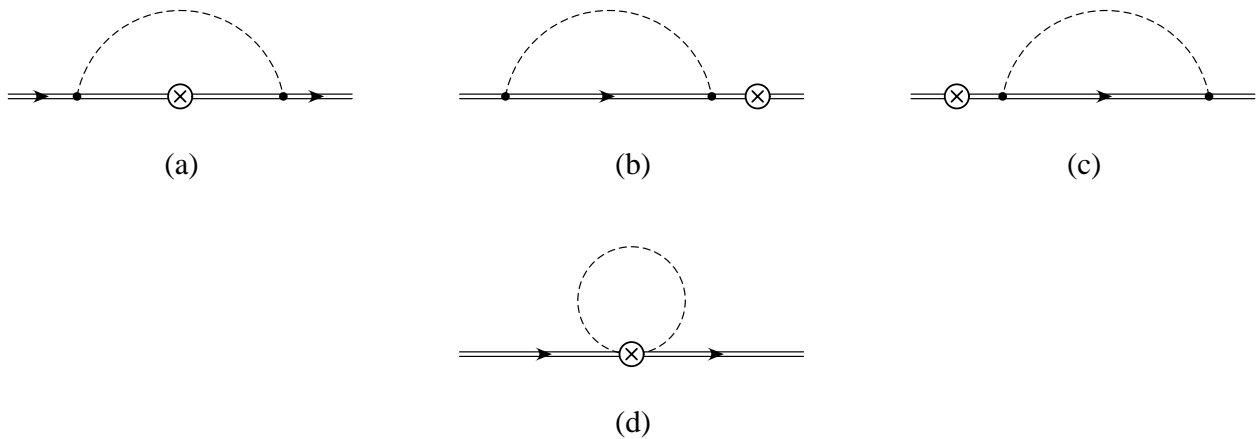}}
\caption{\label{fig:eins}One-loop corrections to the baryon axial vector current.}
\end{figure}

\subsection{One-loop correction: Diagrams \ref{fig:eins}(a,b,c)}

\begin{figure}[ht]
\scalebox{0.4}{\includegraphics{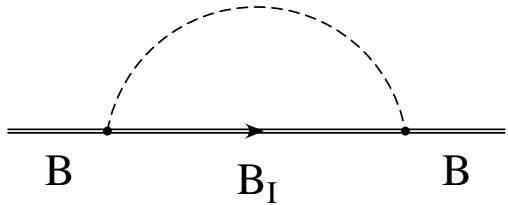}}
\caption{\label{fig:zwei}One-loop wavefunction renormalization graph.}
\end{figure}

We first consider the one-loop wavefunction renormalization graph Fig.~\ref{fig:zwei}, which is part of the diagrams \ref{fig:eins}(b,c).
In this section we restrict ourselves to the computation of the octet meson corrections, such that $\Pi$ denotes $\pi$, $K$, and $\eta$ mesons. In Sec.~\ref{etaPrime}
we will then include the singlet $\eta^\prime$ correction into the analysis.

The Feynman diagram of Fig.~\ref{fig:zwei} depends on the function $F(m_\Pi,\Delta,\mu)$ which is defined by the loop integral
\begin{equation}
\delta^{ij} \, F(m_\Pi,\Delta,\mu) = \frac{i}{f^2} \int \frac{d^4k}{(2\pi)^4} \frac{(\mathbf{k}^i)(-\mathbf{k}^j)} {(k^2-m_\Pi^2)(k\cdot v-\Delta+i\epsilon)}.
\label{eq:Fdef}
\end{equation}
This integral was solved using dimensional regularization in Ref.~\cite{dobogoko}, so $\mu$ in Eq.~(\ref{eq:Fdef}) denotes the scale parameter. Therein, only the leading
nonanalytic pieces were kept explicitly \footnote{Note that in Refs.~\cite{b2,b1}, in dealing with loop integrals, other
regularization schemes were proposed. The central idea in these works was to emphasize the long
distance effects of the integrals and reduce the short distance contributions, which arise from the propagation of Goldstone bosons over distances smaller than a typical
hadronic size. In the present work, however, we will concentrate on the structure of the diagrams themselves and use the former results obtained in dimensional
regularization.}.

The correction arising from the sum of the diagrams of Figs.~\ref{fig:eins}(a,b,c), containing the full dependence on the ratio $\Delta/m_\Pi$, was derived in
Ref.~\cite{fmhjm} and reads
\begin{eqnarray}
\delta A^{kc} & = & \frac12 \left[A^{ja},\left[A^{jb},A^{kc}\right]\right] \Pi_{(1)}^{ab} - \frac12 \left\{ A^{ja}, \left[A^{kc}, \left[\mathcal{M},A^{jb}\right]
\right] \right\} \Pi_{(2)}^{ab} \nonumber \\
&  & \mbox{} + \frac16 \left(\left[A^{ja}, \left[\left[\mathcal{M}, \left[ \mathcal{M},A^{jb}\right]\right],A^{kc}\right] \right] - \frac12
\left[\left[\mathcal{M},A^{ja}\right], \left[\left[\mathcal{M},A^{jb}\right],A^{kc}\right]\right]\right) \Pi_{(3)}^{ab} + \ldots \nonumber \\
\label{eq:dakc}
\end{eqnarray}
Here $\Pi_{(n)}^{ab}$ is a symmetric tensor which contains meson loop integrals with the exchange of a single meson: A meson of flavor $a$ is emitted and a meson
of flavor $b$ is reabsorbed. $\Pi_{(n)}^{ab}$ decomposes into flavor singlet, flavor $\mathbf{8}$, and flavor $\mathbf{27}$ representations as \cite{jen96}
\begin{eqnarray}
\Pi_{(n)}^{ab} = F_\mathbf{1}^{(n)} \delta^{ab} + F_\mathbf{8}^{(n)} d^{ab8} + F_\mathbf{27}^{(n)} \left[
\delta^{a8} \delta^{b8} - \frac18 \delta^{ab} - \frac35 d^{ab8} d^{888}\right], \label{eq:pisym}
\end{eqnarray}
where
\begin{eqnarray}
F_\mathbf{1}^{(n)} & = & \frac18 \left[3F^{(n)}(m_\pi,0,\mu) + 4F^{(n)}(m_K,0,\mu) + F^{(n)}(m_\eta,0,\mu) \right], \label{eq:F1}\\
F_\mathbf{8}^{(n)} & = & \frac{2\sqrt 3}{5} \left[\frac32 F^{(n)}(m_\pi,0,\mu) - F^{(n)}(m_K,0,\mu) - \frac12 F^{(n)}(m_\eta,0,\mu) \right], \label{eq:F8}\\
F_\mathbf{27}^{(n)} & = & \frac13 F^{(n)}(m_\pi,0,\mu) - \frac43 F^{(n)}(m_K,0,\mu) + F^{(n)}(m_\eta,0,\mu). \label{eq:F27}
\end{eqnarray}
Note that Eqs.~(\ref{eq:F1})-(\ref{eq:F27}) are linear combinations of $F^{(n)}(m_\pi,0,\mu)$, $F^{(n)}(m_K,0,\mu)$, and $F^{(n)}(m_\eta,0,\mu)$, where
$F^{(n)}(m_\Pi,0,\mu)$ represents the degeneracy limit
$\Delta/m_\Pi = 0$ of the general function $F^{(n)}(m_\Pi,\Delta,\mu)$, defined as
\begin{equation}
F^{(n)}(m_\Pi,\Delta,\mu) \equiv \frac{\partial^n F(m_\Pi,\Delta,\mu)}{\partial \Delta^n}.
\end{equation}
The first two derivatives of the function read
\begin{eqnarray}
24\pi^2f^2 \, F^{(1)} (m_\Pi,\Delta,\mu) & = & 3\left[\Delta^2-\frac12 m_\Pi^2\right]\ln \frac{m_\Pi^2}{\mu^2}
-6\Delta^2-\frac{11}{2} m_\Pi^2 \nonumber \\
&  & \mbox{} - \left\{ \begin{array}{ll} \displaystyle 3\Delta\sqrt{m_\Pi^2-\Delta^2} \left[\pi-2 \, \textrm{arctan} \left(\frac{\Delta}{\sqrt{m_\Pi^2-\Delta^2}}
\right) \right], & m_\Pi \geq |\Delta| \\[6mm]
\displaystyle 3\Delta\sqrt{\Delta^2-m_\Pi^2} \ln \left[ \frac{\Delta-\sqrt{\Delta^2-m_\Pi^2}}{\Delta+\sqrt{\Delta^2-m_\Pi^2}} \right]. & m_\Pi \leq |\Delta|
\end{array} \right.
\end{eqnarray}
\begin{eqnarray}
24\pi^2f^2 \, F^{(2)} (m_\Pi,\Delta,\mu) & = & 6\Delta \left[\ln\frac{m_\Pi^2}{\mu^2}-1\right] \nonumber \\
&  & \mbox{} - \left\{ \begin{array}{ll} \displaystyle \frac{3(m_\Pi^2-2\Delta^2)}{\sqrt{m_\Pi^2-\Delta^2}} \left[\pi-2 \, \textrm{arctan} \left( \frac{\Delta}
{\sqrt{m_\Pi^2-\Delta^2}} \right) \right], & m_\Pi \geq |\Delta| \\[6mm]
\displaystyle \frac{3(2\Delta^2-m_\Pi^2)}{\sqrt{\Delta^2-m_\Pi^2}} \ln \left[\frac{\Delta-\sqrt{\Delta^2-m_\Pi^2}}{\Delta+\sqrt{\Delta^2-m_\Pi^2}} \right], & m_\Pi
\leq |\Delta| \end{array} \right.
\end{eqnarray}
In the degeneracy limit $\Delta/m_\Pi= 0$ they thus reduce to
\begin{eqnarray}
F^{(1)} (m_\Pi, 0, \mu) & = & - \frac{m_\Pi^2}{16\pi^2f^2} \left(\frac{11}{3} + \ln{\frac{m_\Pi^2}{\mu^2}}\right), \label{eq:fprime} \\
F^{(2)}(m_\Pi, 0, \mu) & = & - \frac{m_\Pi}{8 \pi f^2}. \label{fbprime}
\end{eqnarray}
In Eq.~(\ref{eq:fprime}) the terms involving 11/3 and $\ln(m_\Pi^2/\mu^2)$ are analytic and non-analytic in the quark mass, respectively. The former is scheme dependent and has
the same form as higher dimension terms in the chiral Lagrangian whereas the latter is universal.

For $N_c=3$, the baryon axial vector current $A^{kc}$ has a $1/N_c$ expansion in terms of the four operators of Eq.~(\ref{eq:akc}). The correction $\delta A^{kc}$ --
Eq.~(\ref{eq:dakc}) -- contains $n$-body operators \footnote{An $n$-body operator is one with $n$ $q$'s and $n$
$q^\dagger$'s, namely, it can be written as a polynomial of order $n$ in $J^i$, $T^a$, and $G^{ia}$ \cite{djm95}.}, with $n>N_c$, which are complicated commutators and/or
anticommutators of the one-body operators $J^k$, $T^c$, and $G^{kc}$. All these higher order operators should be reduced and rewritten as linear combinations of the operator
basis, with $n \leq N_c$. The fact that the operator basis is complete and independent facilitates this reduction \cite{djm94,djm95}. In practice, however, dealing with these
expressions becomes rather difficult. Before engaging ourselves in this task, it is convenient to have a useful $1/N_c$ power-counting scheme at hand to save
a considerable effort.

There is a nontrivial $N_c$ dependence of the matrix elements of the generators $J^i$, $T^a$, and $G^{ia}$ in the weight diagrams for the $SU(3)$ flavor
representations of the spin-$\frac12$ and spin-$\frac32$ baryons \cite{djm95}. For instance, factors of $T^a/N_c$ and $G^{ia}/N_c$ are of order 1 somewhere in the weight
diagram, whereas factors of $J^i/N_c$ are of order $1/N_c$ everywhere. If we restrict ourselves to baryons with spins of order unity, the $N_c$ counting rules can
be summarized as \cite{fmhjm}
\begin{equation}
T^a \sim N_c, \qquad G^{ia} \sim N_c, \qquad J^i \sim 1. \label{eq:crules}
\end{equation}
Note that factors of $J^i/N_c$ are $1/N_c$ suppressed relative to factors of $T^a/N_c$ and $G^{ia}/N_c$. Similarly, the meson decay constant $f \propto \sqrt{N_c}$,
so the functions $F^{(n)}(m_\Pi,\Delta,\mu)$ introduce a $1/N_c$ suppression.

In order to evaluate the complicated expressions in Eq.~(\ref{eq:dakc}), the mathematical groundwork developed in Ref.~\cite{djm95} -- which involves a considerable
amount of group theory-- will be used here. First notice that the commutator of an $m$-body operator with an $n$-body operator is an $(m+n-1)$-body operator, namely,
\begin{eqnarray}
\left[\mathcal{O}^{(m)}, \mathcal{O}^{(n)}\right] = \mathcal{O}^{(m+n-1)}. \nonumber
\end{eqnarray}
However, the anticommutator of an $m$-body operator and an $n$-body operator is in general an $(m+n)$-body operator. The $SU(2N_f)$ Lie algebra commutation
relations between one-body operators are given in Table \ref{t:su2fcomm}. Along with these commutation relations, we will use the nontrivial two-body operator
identities for $SU(2N_f)$ quark operators and their transformation properties under $SU(2)\times SU(N_f)$, which were derived in full in Ref.~\cite{djm95}. Let us now
discuss the various terms occurring in the one-loop correction to the baryon axial vector current Eq.~(\ref{eq:dakc}).

\begingroup
\begin{table}
\caption{$SU(2 N_f)$ Commutation relations}
\bigskip
\label{t:su2fcomm}
\centerline{\vbox{ \tabskip=0pt \offinterlineskip
\halign{
\strut\quad $ # $\quad\hfil&\strut\quad $ # $\quad \hfil\cr
\multispan2\hfil $\left[J^i,T^a\right]=0,$ \hfil \cr
\noalign{\medskip}
\left[J^i,J^j\right]=i\epsilon^{ijk} J^k,
&\left[T^a,T^b\right]=i f^{abc} T^c,\cr
\noalign{\medskip}
\left[J^i,G^{ja}\right]=i\epsilon^{ijk} G^{ka},
&\left[T^a,G^{ib}\right]=i f^{abc} G^{ic},\cr
\noalign{\medskip}
\multispan2\hfil$\displaystyle [G^{ia},G^{jb}] = \frac{i}{4}\delta^{ij}
f^{abc} T^c + \frac{i}{2N_f} \delta^{ab} \epsilon^{ijk} J^k + \frac{i}{2} \epsilon^{ijk} d^{abc} G^{kc}.$ \hfill\cr
}}}
\end{table}
\endgroup

\subsubsection{Diagrams \ref{fig:eins}(a,b,c): Degeneracy limit $\Delta/m_{\Pi} = 0$}

The first term in Eq.~(\ref{eq:dakc}) is the double commutator
\begin{eqnarray}
\frac12 \left[A^{ja},\left[A^{jb},A^{kc}\right]\right] \Pi_{(1)}^{ab}, \label{eq:deglim}
\end{eqnarray}
and corresponds to the degeneracy limit $\Delta/m_{\Pi} \! = \! 0$ for the correction to $A^{kc}$. Although this term has been already discussed in the literature
\cite{dm315,djm94,weise,fmhjm}, its explicit computation has not been presented in detail so far.

A crucial observation is the fact that the large-$N_c$ consistency conditions derived in Ref.~\cite{dm315} set this double commutator to be $\mathcal{O}(N_c)$. Naively,
one would expect the double commutator to be $\mathcal{O}(N_c^3)$: one factor of $N_c$ from each $A^{kc}$. However, there are large-$N_c$ cancellations between the Feynman
diagrams of Figs.~\ref{fig:eins}(a,b,c), {\it provided all baryon states in a complete multiplet of the large-$N_c$ $SU(6)$ spin-flavor symmetry are included in the sum
over intermediate states and the axial coupling ratios predicted by this spin-flavor symmetry are used} \cite{fmhjm}. We aim in this section to show explicitly how these
cancellations occur.

For $N_c=3$, it suffices taking the lowest-lying baryon states, which corresponds to the well-known $\mathbf{56}$ dimensional representation of $SU(6)$, namely,
octet and decuplet baryons. For larger $N_c$, there appear more complex representations containing unphysical states with spins greater than $3/2$ and flavor
representations bigger than the $\mathbf{8}$ and $\mathbf{10}$ \cite{weise}. It has already been shown in Ref.~\cite{fmhjm} that the terms $GGG$,
$GG\mathcal{D}_2$, $G\mathcal{D}_2\mathcal{D}_2$, $GG\mathcal{D}_3$, and $GG\mathcal{O}_3$ in the product $AAA$ contribute at the same order to the double
commutator. In the present work we go one step further and also incorporate the terms $\mathcal{D}_2\mathcal{D}_2\mathcal{D}_2$, $G\mathcal{D}_2\mathcal{D}_3$, and
$G\mathcal{D}_2\mathcal{O}_3$ into the analysis. This then means that in the correction to the baryon axial vector current (\ref{eq:deglim}) we will also include terms
that represent $\mathcal{O}(1/N_c^2)$ corrections to the tree-level result $\mathcal{O}(N_c)$. Although our computation will be performed for an arbitrary number of
light quark flavors $N_f$, without loss of generality, in this section we will present our results for the physically interesting case of three light flavors, $N_f=3$.
Results for arbitrary $N_f$ are given in Appendix \ref{app:bo} for completeness.

In order to explicitly show the large-$N_c$ cancellations in Eq.~(\ref{eq:deglim}), it is useful to work out a few examples. At leading order in $N_c$, $A^{kc}$ is
given by $a_1G^{kc}$ so that the double commutator
$[a_1G^{ia},[a_1G^{ib},a_1G^{kc}]]$, for $N_f=3$, yields
\begin{equation}
a_1^3[G^{ia},[G^{ib},G^{kc}]] = \frac{1}{12}a_1^3 \left[3(-f^{bcd}f^{ade}+2d^{bcd}d^{ade})G^{ke} + 4\delta^{bc}G^{ka} + 2d^{abc}J^k\right],
\end{equation}
which is at most $\mathcal{O}(N_c)$ according to the counting rules (\ref{eq:crules}), irrespective of the appropriate contractions of the flavor indices $a$ and
$b$: The contraction with either $\delta^{ab}$, $d^{ab8}$, or $\delta^{a8}\delta^{b8}$ to construct an operator in the flavor singlet, octet, or \textbf{27} representations,
respectively [see Eq.~(\ref{eq:pisym})], does not introduce any additional $N_c$-dependence.

At the next order in the $1/N_c$ expansion one has
\begin{eqnarray}
&  & a_1^2b_2 \frac{1}{N_c} \left([G^{ia},[G^{ib},\mathcal{D}_2^{kc}]] + [G^{ia},[\mathcal{D}_2^{ib},G^{kc}]]
+ [\mathcal{D}_2^{ia},[G^{ib},G^{kc}]]\right) = a_1^2b_2 \frac{1}{N_c} \Bigl(\nonumber \\
&  & \mbox{} \frac16 if^{abc}J^k - \frac54 f^{bcd}f^{ade}\mathcal{D}_2^{ke} + \frac13 \delta^{ab} \mathcal{D}_2^{kc}
+ \frac13 \delta^{ac}\mathcal{D}_2^{kb} + \frac13 \delta^{bc} \mathcal{D}_2^{ka} \nonumber \\
&  & \mbox{} + d^{abe}G^{ke}T^c+d^{bce}G^{ke}T^a+d^{ace}G^{ke}T^b + \frac12 \epsilon^{kim} (f^{aed}d^{bce}+d^{aed}f^{bce})J^iG^{md} \nonumber \\
&  & \mbox{} +\epsilon^{kim}(f^{bcd}G^{ma} + f^{acd}G^{mb} + f^{abd}G^{mc})G^{id} \Bigr),
\end{eqnarray}
which is also at most $\mathcal{O}(N_c)$. As for the additional subleading terms, the calculation is straightforward although tedious in practice in view of the
considerable amount of group theory involved. The explicit expressions for arbitrary $N_c$ and $N_f$ may be found in Appendix \ref{app:bo}. To the order of approximation
adopted here, the different flavor contributions originating from diagrams ~\ref{fig:eins}(a,b,c), in the degeneracy limit, can be organized as follows
\footnote{Note that, in this section, we restrict ourselves to operators that have nonzero matrix elements between \textit{octet} baryons -- the full expressions can be
found in Appendix \ref{app:bo}.}

\begin{enumerate}
\item \textit{Flavor singlet contribution}
\begin{eqnarray}
&  & [\,A^{ia}, [\,A^{ia},A^{kc}\,]\,] = \left[\frac{23}{12} a_1^3 - \frac{2(N_c+3)}{3N_c}a_1^2b_2 + \frac{N_c^2+6N_c-54}{6N_c^2} a_1b_2^2 \right. \nonumber \\
&  & \mbox{} \left. - \frac{N_c^2+6N_c+2}{N_c^2} a_1^2b_3 - \frac{N_c^2+6N_c-3}{N_c^2} a_1^2c_3 - \frac{12(N_c+3)}{N_c^3} a_1b_2b_3\right] G^{kc} \nonumber \\
&  & \mbox{} + \frac{1}{N_c} \left[\frac{101}{12} a_1^2b_2 + \frac{4(N_c+3)}{3N_c} a_1b_2^2 - \frac{3(N_c+3)}{N_c} a_1^2b_3  - \frac{N_c+3}{2N_c} a_1^2c_3 \right.
\nonumber \\
&  & \mbox{} \left. + \frac{N_c^2+6N_c-18}{6N_c^2} b_2^3 + \frac{N_c^2+6N_c+2}{N_c^2} a_1b_2b_3 - \frac{3(N_c^2+6N_c-24)}{2N_c^2}a_1b_2c_3 \right] \mathcal{D}_2^{kc}
\nonumber \\
&  & \mbox{} + \frac{1}{N_c^2} \left[\frac{11}{4} a_1 b_2^2 + \frac{51}{4} a_1^2b_3 + 2a_1^2c_3 + \frac{17(N_c+3)}{3N_c} a_1b_2b_3 - \frac{9(N_c+3)}{2N_c}a_1b_2c_3
\right] \mathcal{D}_3^{kc}\nonumber \\
&  & \mbox{} + \frac{1}{N_c^3} \left[\frac52 b_2^3+\frac{11}{3}a_1b_2b_3 + 19 a_1b_2c_3 \right] \mathcal{D}_4^{kc} +\mathcal{O}(G\mathcal{D}_3\mathcal{D}_3).
\label{eq:fsinglet}
\end{eqnarray}

The symbol $\mathcal{O}(G\mathcal{D}_3\mathcal{D}_3)$ means that, in the double commutator structure $AAA$, we have included all terms up to six-body operators, such as
$G\mathcal{D}_2\mathcal{D}_3$, but have neglected contributions which are seven-body operators -- like $G\mathcal{D}_3\mathcal{D}_3$ -- or higher.

\item \textit{Flavor octet contribution}
\begin{eqnarray}
&  & d^{ab8} [\,A^{ia},[\,A^{ib},A^{kc}\,]\,] = \nonumber \\
&  & \mbox{} \left[\frac{11}{24} a_1^3 - \frac{2(N_c+3)}{3N_c} a_1^2b_2 - \frac{9}{2 N_c^2}a_1b_2^2 - \frac{5}{N_c^2} a_1^2b_3 + \frac{3}{2 N_c^2} a_1^2c_3
-\frac{6(N_c+3)}{N_c^3} a_1b_2b_3\right] d^{c8e}G^{ke} \nonumber \\
&  & \mbox{} + \frac{1}{8N_c} \left[23 a_1^2b_2 - \frac{2(N_c+3)}{N_c}(6a_1^2b_3 + a_1^2c_3) - \frac{12}{N_c^2}(b_2^3+2a_1b_2b_3-12a_1b_2c_3) \right] d^{ce8}
\mathcal{D}_2^{ke} \nonumber \\
&  & \mbox{} - \frac{1}{6N_c} \left[4a_1^2b_2 + \frac{N_c+3}{N_c} \left(a_1b_2^2 + 6a_1^2b_3 + 6a_1^2c_3\right) + \frac{36}{N_c^2}a_1b_2b_3 \right] \{G^{kc},T^8\}
\nonumber \\
&  & \mbox{} + \frac{1}{6N_c} \left[11a_1^2b_2 + \frac{2(N_c+3)}{N_c} a_1b_2^2 + \frac{48}{N_c^2}a_1b_2b_3 \right] \{G^{k8},T^c\} \nonumber \\
&  & \mbox{} + \frac{1}{24N_c^2} \left[27 a_1b_2^2 + 65a_1^2b_3 + 8a_1^2c_3 + \frac{36(N_c+3)}{N_c} a_1b_2b_3 - \frac{46(N_c+3)}{N_c} a_1b_2c_3\right] d^{c8e}
\mathcal{D}_3^{ke} \nonumber \\
&  & \mbox{} + \frac{1}{6N_c^2} \left[3a_1b_2^2 - 2a_1^2b_3 + 30a_1^2c_3 + \frac{4(N_c+3)}{N_c} a_1b_2b_3 \right] \{G^{kc},\{J^r,G^{r8}\}\} \nonumber \\
&  & \mbox{} + \frac{1}{6N_c^2} \left[3a_1b_2^2 + 28a_1^2b_3 - 15a_1^2c_3 + \frac{4(N_c+3)}{N_c} a_1b_2b_3 \right] \{G^{k8},\{J^r,G^{rc}\}\} \nonumber \\
&  & \mbox{} + \frac{1}{3N_c^2} \left[12a_1^2b_3 - 2a_1^2c_3 -\frac{2(N_c+3)}{N_c} a_1b_2c_3 \right] \{J^k,\{G^{rc},G^{r8}\}\} \nonumber \\
&  & \mbox{} + \frac{1}{12N_c^2} \left[2a_1b_2^2 - 9a_1^2b_3 - \frac32 a_1^2c_3
-\frac{N_c+3}{N_c}\left(b_2^3 - 6a_1b_2b_3 + 9a_1b_2c_3\right) \right] \{J^k,\{T^c,T^8\}\} \nonumber \\
&  & \mbox{} + \frac{1}{3N_c^3}\left(3b_2^3-21a_1b_2b_3+20a_1b_2c_3\right) \{\mathcal{D}_2^{kc},\{J^r,G^{r8}\}\} \nonumber \\
&  & \mbox{} + \frac{1}{6N_c^3}\left(24a_1b_2b_3-23a_1b_2c_3\right) \{\mathcal{D}_2^{k8},\{J^r,G^{rc}\}\} \nonumber \\
&  & \mbox{} + \frac{1}{3N_c^3} \left( a_1b_2b_3-2a_1b_2c_3\right) \{J^2,\{G^{kc},T^8\}\}
+ \frac{1}{6N_c^3} \left(20a_1b_2b_3+11a_1b_2c_3\right) \{J^2,\{G^{k8},T^c\}\} \nonumber \\
&  & \mbox{} + \frac{1}{128N_c^3} \left(10a_1b_2b_3+71a_1b_2c_3\right) \left(\{J^2,[G^{k8},\{J^r,G^{rc}\}]\}
- \{J^2,[G^{kc},\{J^r,G^{r8}\}]\} \right. \nonumber \\
&  & \mbox{} + \left. \{J^k,[\{J^m,G^{mc}\},\{J^r,G^{r8}\}]\} \right) + \frac{1}{4N_c^3} \left(3b_2^3+ 6a_1b_2b_3+24a_1b_2c_3 \right) d^{c8e} \mathcal{D}_4^{ke}+
\mathcal{O}(G\mathcal{D}_3\mathcal{D}_3).
\nonumber \\
\label{eq:foctet}
\end{eqnarray}

\item \textit{Flavor $\mathbf{27}$ contribution}
\begin{eqnarray}
&  & [\,A^{i8}, [\,A^{i8},A^{kc}\,]\,] = \nonumber \\
&  & \mbox{} \left[\left(\frac14 a_1^3 - \frac{1}{N_c^2} \left(2a_1b_2^2+2a_1^2b_3-a_1^2c_3 \right) \right)
f^{c8e} f^{8eg} + \frac12 \left(a_1^3 + \frac{1}{N_c^2} \left(2a_1^2b_3-a_1^2c_3 \right) \right) d^{c8e}d^{8eg} \right] G^{kg} \nonumber \\
&  & \mbox{} + \frac{1}{N_c} \left[\frac{1}{12}a_1^2b_2 \left(4\delta^{cg} + 21f^{c8e}f^{8eg}\right) + \frac{1}{N_c^2} (-b_2^3+9a_1b_2c_3) f^{c8e}f^{8eg} \right]
\mathcal{D}_2^{kg}
\nonumber \\
&  & \mbox{} + \frac{1}{2N_c} a_1^2b_2 \left(2d^{c8e} \{G^{ke}, T^8\} + d^{88e} \{G^{ke}, T^c\}\right)
+ \frac{1}{N_c} \left(a_1^2b_2+\frac{4}{N_c^2}a_1b_2b_3\right)i f^{c8e} [G^{k8},\{J^r,G^{re}\}]
\nonumber \\
&  & \mbox{} -\frac{4}{N_c^3}a_1b_2b_3i f^{c8e} [G^{ke},\{J^r,G^{r8}\}] + \frac{1}{N_c^2} a_1^2c_3 d^{88e} \{G^{kc},\{J^r,G^{re}\}\} + \frac{1}{2N_c^3} b_2^3 \{\mathcal{D}_2^{kc},\{T^8,T^8\}\} \nonumber \\
&  & \mbox{} + \frac{1}{12N_c^2} \left[9a_1b_2^2 f^{c8e}f^{8eg} + a_1^2b_3 \left(8\delta^{cg} + 9f^{c8e}f^{8eg} + 6d^{c8e} d^{8eg}\right) + 6 a_1^2c_3 d^{c8e} d^{8eg}
\right] \mathcal{D}_3^{kg} \nonumber \\
&  & \mbox{}
+ \frac{1}{2N_c^3} \left(-2a_1b_2b_3 + a_1b_2c_3\right) \left( 2d^{c8e}\{\mathcal{D}_2^{k8},\{J^r,G^{re}\}\} +
d^{88e} \{\mathcal{D}_2^{kc},\{J^r,G^{re}\}\} \right)
\nonumber \\
&  & \mbox{} + \frac{1}{2N_c^2} a_1b_2^2 \left(\{G^{kc},\{T^8,T^8\}\} + 2\{G^{k8},\{T^c,T^8\}\} \right) + \frac{1}{N_c^2} (4 a_1^2b_3 - a_1^2c_3) d^{c8e}
\{G^{ke},\{J^r,G^{r8}\}\} \nonumber \\
&  & \mbox{} - \frac{1}{2N_c^2} \left(6a_1^2b_3 + a_1^2c_3\right) d^{c8e} \{J^k,\{G^{re},G^{r8}\}\} + \frac{1}{N_c^2} \left(2 a_1^2b_3 - a_1^2c_3\right)
\{G^{rc},\{G^{r8},G^{k8}\}\} \nonumber \\
&  & \mbox{} - \frac{1}{N_c^2} (2 a_1^2b_3 + a_1^2c_3) \{G^{kc},\{G^{r8},G^{r8}\}\} + \frac{1}{2N_c^2} \left(-2a_1^2b_3 + 3a_1^2c_3\right)  d^{c8e}
\{G^{k8},\{J^r,G^{re}\}\} \nonumber \\
&  & \mbox{} + \frac{1}{2N_c^2} \left(2a_1^2b_3 - a_1^2c_3\right) \left(d^{88e} \{J^k,\{G^{rc},G^{re}\}\} + d^{88e} \{G^{ke},\{J^r,G^{rc}\}\} \right) \nonumber \\
&  & \mbox{} + \frac{2}{N_c^3} a_1b_2b_3 \left( \{\{J^r,G^{r8}\},\{G^{k8},T^c\}\}
+ \{\{J^r,G^{r8}\},\{G^{kc},T^8\}\} + \{\{J^r,G^{rc}\},\{G^{k8},T^8\}\} \right) \nonumber \\
&  & \mbox{} -\frac{2}{N_c^3}a_1b_2c_3\left(2\{\mathcal{D}_2^{k8},\{G^{rc},G^{r8}\}\} + \{\mathcal{D}_2^{kc},\{G^{r8},G^{r8}\}\} \right) \nonumber \\
&  & \mbox{} + \frac{1}{2 N_c^3} \left(2a_1b_2b_3+a_1b_2c_3\right)\left( d^{88e} \{J^2,\{G^{ke},T^c\}\} + 2 d^{c8e}\{J^2,\{G^{ke},T^8\}\} \right) \nonumber \\
&  & \mbox{} + \frac{1}{N_c^3}\left[\frac23 a_1b_2c_3 \delta^{cg} + \frac12 \left(b_2^3+9a_1b_2c_3 \right) f^{c8e}f^{8eg}  \right] \mathcal{D}_4^{kg}
+ \frac{2}{N_c^3}a_1b_2c_3 if^{c8e}\{J^2,[G^{k8},\{J^r,G^{re}\}]\}\nonumber \\
&  & \mbox{}
+ \frac{1}{N_c^3} \left(2a_1b_2b_3-a_1b_2c_3\right) if^{c8e}\{J^2,[G^{ke},\{J^r,G^{r8}\}]\} \nonumber \\
&  & \mbox{} + \frac{2}{N_c^3} \left(a_1b_2b_3-a_1b_2c_3\right) i f^{c8e} \{J^k,[\{J^i,G^{ie}\},\{J^r,G^{r8}\}]\}
+
\mathcal{O}(G\mathcal{D}_3\mathcal{D}_3). \label{eq:f27}
\end{eqnarray}

\end{enumerate}

In order for Eq.~(\ref{eq:f27}) to be a truly $\mathbf{27}$ contribution it is understood that flavor singlet and octet contributions should be subtracted off from
this equation. For computational purposes the one-body operators $T^8$ and $G^{i8}$ can be written in terms of the strange quark number operator $N_s$ and the
strange quark spin operator $J_s^i$ as \cite{djm95}
\begin{eqnarray}
T^8 & = & \frac{1}{2 \sqrt 3} (N_c - 3 N_s), \\
G^{i8} & = & \frac{1}{2 \sqrt 3} (J^i - 3 J_s^i).
\end{eqnarray}
These operators are order $N_c$ and order 1, respectively.

Equations (\ref{eq:fsinglet})--(\ref{eq:f27}) have been rearranged to display leading and subleading terms in $1/N_c$ explicitly. Notice that only baryon operators
with nonvanishing matrix elements between octet baryons have been kept in these equations (for the full expressions see Appendix \ref{app:bo}). Although the resulting
expressions are rather lengthy, they are indeed enlightening. It is now evident that large-$N_c$ cancellations occur in the evaluation of the double commutator so
that it is at most $\mathcal{O}(N_c)$, according to the counting rules (\ref{eq:crules}). The loop integrals are inversely proportional to $f^2$, which introduces
a $1/N_c$ suppression. Therefore the net one-loop correction Eq.~(\ref{eq:deglim}) is $\mathcal{O}(1)$, or $1/N_c$ times the tree-level value, which is
$\mathcal{O}(N_c)$. The large-$N_c$ cancellations in the renormalization of the baryon axial vector current to one-loop have thus been explicitly shown to occur in
the degeneracy limit $\Delta/m_{\Pi} = 0$.

\subsubsection{Diagrams \ref{fig:eins}(a,b,c): Nondegeneracy case $\Delta/m_{\Pi} \neq 0$}

Let us now discuss the additional terms that contribute to the renormalization of the baryon axial vector current for finite $\Delta/m_\Pi$. The procedure for
obtaining these terms is discussed in Ref.~\cite{fmhjm}. Specifically, let us consider the second term in Eq.~(\ref{eq:dakc}),
\begin{eqnarray}
\frac12 \left\{A^{ja},\left[A^{kc}, \left[\mathcal{M},A^{jb}\right]\right] \right\} \Pi_{(2)}^{ab}. \label{eq:sterm}
\end{eqnarray}
This expression contains one insertion of the baryon mass matrix $\mathcal{M}$ introduced in Eq.~(\ref{eq:mop}) and thus represents the leading term in the nondegenerate
case. The large-$N_c$ counting rules imply that multiple insertions of the $J^2$ factor in $\mathcal{M}$ constitute the dominant $1/N_c$ corrections from the baryon mass
splittings: In Ref.~\cite{fmhjm}, it has been shown that one insertion of $J^4$ in the term linear in $\mathcal{M}$ is $1/N_c$ suppressed relative to two insertions of
$J^2$ in the term quadratic in $\mathcal{M}$ -- the third term in Eq.~(\ref{eq:dakc}). Moreover, in the same reference it was concluded that the quantities $GGGJ^2$ and
$GG\mathcal{D}_2J^2$ in the product $AAA{\mathcal M}$ contribute at the same order in Eq.~(\ref{eq:sterm}) and should be retained in the series Eq.~(\ref{eq:dakc}).

Returning to Eq.~(\ref{eq:sterm}), a new large-$N_c$ cancellation for the specific commutator-anticommutator structure $GGGJ^2$ was found in Ref.~\cite{fmhjm}. Naively,
one would expect this contribution to be of $\mathcal{O}(N_c^3)$: The two operators $J$ may be eliminated with the two commutators, such that we are left with a product of
three operators $GGG$, each one contributing a factor of $N_c$. However, the explicit calculation of the singlet contribution of the operator expression $GGGJ^2$ shows that
it is of $\mathcal{O}(N_c^2)$, i.e., suppressed by one factor of $N_c$. We would like to see whether the same pattern repeats itself in the octet and the $\mathbf{27}$
piece of $GGGJ^2$, and whether new large-$N_c$ cancellations also occur in the operator structure $GG\mathcal{D}_2J^2$. The expressions, when retaining both structures
$GGGJ^2$ and $GG\mathcal{D}_2J^2$ in the product $AAA{\mathcal M}$, read:

\begin{enumerate}

\item \textit{Flavor singlet contribution}
\begin{eqnarray}
&  & \left\{A^{ja}, \left[A^{kc}, \left[\mathcal{M},A^{ja}\right] \right] \right\} = \nonumber \\
&  & \mbox{} \frac{m_2}{2 N_c} \left\{\left[-a_1^3 + \frac{4(N_c+3)}{N_c}
 a_1^2b_2\right] G^{kc} + \left[(N_c+3)a_1^3 + \frac{N_c^2 + 6N_c-29}{N_c}a_1^2b_2\right] \mathcal{D}_2^{kc} \right. \nonumber \\
&  & \mbox{} + \left. \left[-a_1^3 + \frac{N_c+3}{N_c}a_1^2b_2\right] \mathcal{D}_3^{kc} - \frac{4}{N_c} a_1^2b_2 \mathcal{D}_4^{kc} \right\} + \dots \label{eq:ms}
\end{eqnarray}

\item \textit{Flavor octet contribution}
\begin{eqnarray}
&  & d^{ab8} \left\{A^{ja}, \left[A^{kc}, \left[\mathcal{M},A^{jb} \right]\right] \right\} =  \frac{m_2}{N_c} \left\{\left[\frac14 a_1^3 +\frac{N_c+3}{N_c} a_1^2b_2\right] d^{c8e} G^{ke} \right. \nonumber \\
&  & \mbox{} + \frac14 \left[(N_c+3) a_1^3 - \frac{25}{N_c} a_1^2b_2 \right] d^{c8e} \mathcal{D}_2^{ke} - \frac14 \left[a_1^3 - \frac{N_c+3}{N_c} a_1^2b_2 \right] d^{c8e} \mathcal{D}_3^{ke} \nonumber \\
&  & \mbox{} - \frac12 a_1^3 \{G^{kc},\{J^r,G^{r8}\}\} + \frac18 \left[a_1^3 +  \frac{2(N_c+3)}{N_c}a_1^2b_2 \right] \{J^k,\{T^c,T^8\}\} \nonumber \\
&  & \mbox{} - \frac16 \left[2 a_1^3 - \frac{N_c+3}{N_c} a_1^2b_2\right] \left( \{J^k,\{G^{rc},G^{r8}\}\} - \{G^{k8},\{J^r,G^{rc}\}\} \right) \nonumber \\
&  & \mbox{} + \frac{1}{N_c} a_1^2b_2 \left[-\frac32\{T^c,G^{k8}\}+\{G^{kc},T^8\} - \frac12 d^{c8e} \mathcal{D}_4^{ke}
+ \frac12 \{\mathcal{D}_2^{k8},\{J^r,G^{rc}\}\} \right. \nonumber \\
&  & \mbox{} \left. \left. - \frac43 \{\mathcal{D}_2^{kc},\{J^r,G^{r8}\}\} - \frac16 \{J^2,\{G^{k8},T^c\}\} \right] \right\} + \ldots
\label{eq:mo}
\end{eqnarray}
\item \textit{Flavor $\mathbf{27}$ contribution}
\begin{eqnarray}
&  & \left\{A^{j8}, \left[A^{kc}, \left[\mathcal{M},A^{j8}\right] \right] \right\} = \nonumber \\
&  & \mbox{} \frac{m_2}{N_c} \left\{a_1^3 \left[
\frac13 G^{kc} - \frac12 \left(
d^{c8e} d^{e8d} - d^{ced} d^{e88} - f^{c8e} f^{e8d} \right) G^{kd} - \frac12 d^{c8e} \{G^{k8},\{J^r,G^{re}\}\}  \right. \right. \nonumber \\
&   & \mbox{} \left. + \frac12 d^{c8e} \{J^k,\{G^{re},G^{r8}\}\} \right]
+ \frac{1}{N_c} a_1^2b_2 \left[
-\frac{15}{4}f^{c8e}f^{8eg}\mathcal{D}_2^{kg} + \frac{i}{2}f^{c8e}[G^{ke},\{J^r,G^{r8}\}] \right. \nonumber \\
&  & \mbox{} - if^{c8e}[G^{k8},\{J^r,G^{re}\}] - \frac12 f^{c8e}f^{8eg}\mathcal{D}_4^{kg}  + \{\mathcal{D}_2^{kc},\{G^{r8},G^{r8}\}\}
 + \{\mathcal{D}_2^{k8},\{G^{rc},G^{r8}\}\}  \nonumber \\
&  & \mbox{} - \frac12 \{\{J^r,G^{rc}\},\{G^{k8},T^8\}\} - \frac12 \{\{J^r,G^{r8}\},\{G^{k8},T^c\}\} \nonumber \\
&  & \mbox{} \left. \left. + \frac{i}{2}f^{c8e}\{J^k,[\{J^i,G^{ie}\},\{J^r,G^{r8}\}]\} \right] \right\} + \dots \label{eq:mt}
\end{eqnarray}
\end{enumerate}
Again, in  Eq.~(\ref{eq:mt}), the singlet and octet pieces need be subtracted off in order to have a purely \textbf{27} contribution.

First of all, as can be seen in Appendix \ref{app:bo}, the new cancellation observed in the singlet piece of $GGGJ^2$ indeed repeats itself in the octet and the
$\mathbf{27}$: the three expressions (\ref{GJ2GG-singlet}), (\ref{GJ2GG-octet}) and (\ref{GJ2GG-27}) are indeed of $\mathcal{O}(N_c^2)$. Furthermore, it is evident
from the expressions (\ref{GJ2GD2-singlet}), (\ref{GJ2GD2-octet}) and (\ref{GJ2GD2-27}) in the same Appendix, that the new large-$N_c$ cancellation identified in $GGGJ^2$
does not occur in $GG\mathcal{D}_2J^2$. As one would expect, $GG\mathcal{D}_2J^2$ is of $\mathcal{O}(N_c^3)$: Eliminating two $J$'s with the two commutators, one is
left with the operator product $GGJT$, which is $\mathcal{O}(N_c^3)$, according to the counting rules (\ref{eq:crules}).

This then means that the correction to $\delta A^{kc}$ originating from  Eq.~(\ref{eq:sterm}) is $\mathcal{O}(1)$ and thus consistent with being a quantum correction:
Naively, one would expect the operator expression $\left\{A^{ja},\left[A^{kc}, \left[\mathcal{M}, A^{jb}\right] \right]\right\}$ to be $\mathcal{O}(N_c^2)$ so that
the correction Eq.~(\ref{eq:sterm}) would be $\mathcal{O}(N_c)$, since $f \propto \sqrt{N_c}$. However, a close inspection of Eqs.~(\ref{eq:ms})-(\ref{eq:mt})
reveals that these equations exhibit at most a linear dependence in $N_c$, \textit{i.e.}, large-$N_c$ cancellations occur in the structure of the operator factor
in such a way that it is at most $\mathcal{O}(N_c)$. Therefore, the correction Eq.~(\ref{eq:sterm}) is $\mathcal{O}(1)$, or $1/N_c$ times the tree level value and
contributes to the same order as Eq.~(\ref{eq:deglim}). The general structure of these cancellations was analyzed in Ref.~\cite{fmhjm} and has been shown explicitly here.

Finally, there are the two remaining terms in Eq.~(\ref{eq:dakc}) with two mass insertions,
\begin{equation}
\label{mmAAA}
\frac16 \left(\left[A^{ja}, \left[\left[J^2, \left[ J^2,A^{jb}\right]\right],A^{kc}\right] \right] - \frac12
\left[\left[J^2,A^{ja}\right], \left[\left[J^2,A^{jb}\right],A^{kc}\right]\right]\right) \Pi_{(3)}^{ab},
\end{equation}
which are both of $\mathcal{O}(N_c^3)$: eliminating the four $J$'s with the four commutators, we are left with three $G$'s, each one contributing a factor of $N_c$,
according to the counting rules (\ref{eq:crules}). Interestingly, as shown below for the singlet contribution, there is a new large-$N_c$ cancellation in the first term
of Eq.~(\ref{mmAAA}):
\begin{equation}
\left[G^{ia}, \left[\left[J^2, \left[ J^2,G^{ia}\right]\right],G^{kc}\right] \right] = -\frac32 (N_c+3) \mathcal{D}_2^{kc} + 2 \mathcal{D}_3^{kc} + 3 \mathcal{O}_3^{kc}.
\end{equation}
The right hand side is at most of $\mathcal{O}(N_c^2)$: The order $N_c^3$ part vanishes. We have checked that the same pattern repeats itself in the octet and the
$\mathbf{27}$ piece -- the explicit expressions will be given elsewhere.

As for the second term in Eq.~(\ref{mmAAA}), there is no new cancellation as can be seen in the singlet piece
\begin{equation}
\left[\left[J^2,G^{ia}\right], \left[\left[J^2,G^{ia}\right],G^{kc}\right]\right] = \left[-N_c (N_c+6) + 3 \right] G^{kc} + \frac52 (N_c+3) \mathcal{D}_2^{kc}
- 2 \mathcal{D}_3^{kc} - 2 \mathcal{O}_3^{kc},
\end{equation}
where the right hand side is of $\mathcal{O}(N_c^3)$, as one would naively expect. The octet and $\mathbf{27}$ pieces are of the same order $N_c^3$.

\subsection{One-loop correction: Diagram \ref{fig:eins}(d)}

The one-loop correction to the baryon axial vector current from the diagram of Fig.~\ref{fig:eins}(d) is given by the expression
\begin{eqnarray}
\label{corr1d}
\delta A^{kc} = - \frac12 \left[T^a, \left[T^b,A^{kc}\right]\right] \Pi^{ab}, \label{eq:loopd}
\end{eqnarray}
where $\Pi^{ab}$ is a symmetric tensor with a structure similar to the one introduced in Eq.~(\ref{eq:pisym}), namely,
\begin{eqnarray}
\Pi^{ab} = I_\mathbf{1} \delta^{ab} + I_\mathbf{8} d^{ab8} + I_\mathbf{27} \left[\delta^{a8} \delta^{b8} - \frac18 \delta^{ab} - \frac35 d^{ab8} d^{888}\right].
\label{eq:psymd}
\end{eqnarray}
Again, the flavor singlet, octet, and $\mathbf{27}$ tensors in Eq.~(\ref{eq:psymd}) are proportional to flavor singlet $I_\mathbf{1}$, flavor octet
$I_\mathbf{8}$, and flavor $\mathbf{27}$ $I_\mathbf{27}$ linear combinations of the loop integrals $I(m_\pi,\mu)$, $I(m_K,\mu)$, and $I(m_\eta,\mu)$, reading
\begin{eqnarray}
I(m_\Pi,\mu) & = & \frac{i}{f^2} \int \frac{d^4k}{{(2\pi)}^4} \frac{1}{k^2 - m_\Pi^2} \nonumber \\
& = & \mbox{} \frac{m_\Pi^2}{16 \pi^2 f^2} \left[ \ln{ \frac{m_\Pi^2}{\mu^2}} - 1\right]. \label{eq:intd}
\end{eqnarray}
They enter the linear combinations as
\begin{eqnarray}
I_\mathbf{1} & = & \frac18 \left[3I(m_\pi,\mu) + 4I(m_K,\mu) + I(m_\eta,\mu)\right], \label{eq:I1}\\
I_\mathbf{8} & = & \frac{2\sqrt 3}{5} \left[\frac32 I(m_\pi,\mu) - I(m_K,\mu) - \frac12 I(m_\eta,\mu) \right], \label{eq:I8} \\
I_\mathbf{27} & = & \frac13 I(m_\pi,\mu) - \frac43 I(m_K,\mu) + I(m_\eta,\mu). \label{eq:I27}
\end{eqnarray}

A straightforward computation yields the following flavor contributions for $N_f=3$:
\begin{enumerate}
\item \textit{Flavor singlet contribution}
\begin{eqnarray}
[T^a,[T^a,A^{kc}]] = 3 A^{kc}, \label{eq:sind}
\end{eqnarray}
\item \textit{Flavor octet contribution}
\begin{eqnarray}
d^{ab8} [T^a,[T^b,A^{kc}]] = \frac32 d^{c8e} A^{ke}, \label{eq:octd}
\end{eqnarray}
\item \textit{Flavor $\mathbf{27}$ contribution}
\begin{eqnarray}
[T^8,[T^8,A^{kc}]] = f^{c8e} f^{8eg} A^{kg}. \label{eq:27d}
\end{eqnarray}
\end{enumerate}

The double commutators in Eqs.~(\ref{eq:sind})-(\ref{eq:27d}) are proportional to $A^{kc}$ so they are $\mathcal{O}(N_c)$; thus the one-loop correction of
Fig~\ref{fig:eins}(d) is at most $\mathcal{O}(1)$ since $f^2$ scales like $N_c$. Consequently, this correction is of the same order as the one arising from the
sum of Figs.~\ref{fig:eins}(a,b,c), i.e., it is of order $1/N_c$ relative to the tree-level contribution and does not involve any cancellations between octet and decuplet
states.

\subsection{Total one-loop correction in the degeneracy limit $\Delta/m_\Pi = 0$}

In the limit $\Delta/m_\Pi= 0$ the one-loop correction to $A^{kc}$ becomes
\begin{equation}
\delta A^{kc} = \frac12 \left[A^{ja},\left[A^{jb},A^{kc}\right]\right] \Pi_{(1)}^{ab} - \frac12 \left[T^a, \left[T^b,A^{kc}\right]\right] \Pi^{ab}. \label{eq:axdeg}
\end{equation}

The matrix elements between spin-$\frac12$ baryon states of the space components of the renormalized baryon axial vector current, $A^{kc}+\delta A^{kc}$, are discussed in
detail in Appendix \ref{app:mtx}. These matrix elements yield the coupling constants $g_A$. Our interest in computing these quantities relies on the fact that our
calculations can be compared with results obtained within other approaches. Specifically, a direct comparison can be carried out with $g_A$ obtained within the
framework of heavy baryon chiral perturbation theory originally introduced in Refs.~\cite{jm255,jm259}. In these references the calculation was performed assuming
$m_u\!=\!m_d\!=\!0$ and vanishing decuplet-octet mass difference. In the next section we shall redo the calculation for arbitrary quark
masses \footnote{Of course, the quark masses have still to be small with respect to the scale of chiral symmetry breaking.}.
This will allow us to identify individual contributions of $\pi$, $K$, and $\eta$ mesons in the loops.

\section{The baryon axial vector current in heavy baryon chiral perturbation theory}
\label{HBCHPT}

The heavy baryon chiral Lagrangian was constructed \cite{jm255,jm259} in terms of the octet meson field, the baryon octet $B_v$, and the baryon decuplet $T^\mu_{abc}$
fields. The lowest order Lagrangian is given by
\begin{eqnarray}
\mathcal{L}_{\text{baryon}} & = & i \, \text{Tr} \, \bar{B}_v (v \cdot \mathcal{D}) B_v  - i \, \bar{T}_v^\mu (v \cdot \mathcal{D}) T_{v\mu} + \Delta \,
\bar{T}_v^\mu T_{v\mu} + 2 \, D \, \text{Tr} \, \bar{B}_v S_v^\mu \{\mathcal{A}_\mu, B_v\} \nonumber \\
&  & \mbox{} + 2 \, F \, \text{Tr} \, \bar{B}_v S_v^\mu [ \mathcal{A}_\mu, B_v ] + \mathcal{C} \, ( \bar{T}_v^\mu \mathcal{A}_\mu B_v + \bar{B}_v \mathcal{A}_\mu T_v^\mu) + 2 \, \mathcal{H} \,
\bar{T}_v^\mu S_v^\nu \mathcal{A}_\nu T_{v\mu}, \label{eq:efflag}
\end{eqnarray}
where $D$, $F$, $\mathcal C$, and $\mathcal H$ are the baryon-pion couplings and $\Delta$ is the decuplet-octet mass difference as defined in the preceding sections.

\subsection{Chiral corrections to the baryon axial vector current}

The one-loop corrections to the axial vector current arise from the Feynman graphs displayed in Fig.~\ref{fig:eins}. The renormalized current 
\footnote{In this section we use the notation and conventions of Refs.~\cite{jm255,jm259}.} can be written as
\begin{eqnarray}
\langle B_j | J_\mu^A | B_i \rangle & = & \left[\alpha_{B_jB_i} - \sum_\Pi \left(\bar{\beta}_{B_jB_i}^\Pi - \bar{\lambda}_{B_jB_i}^\Pi \alpha_{B_jB_i}\right)
F^{(1)}(m_\Pi,0,\mu) + \sum_\Pi \gamma_{B_jB_i}^\Pi I(m_\Pi,\mu) \right]\nonumber \\
&  & \times \bar{u}_{B_j} \gamma_\mu \gamma_5 u_{B_i},\label{eq:axren}
\end{eqnarray}
where $\alpha_{B_jB_i}$ is the tree-level result, $\bar{\beta}_{B_jB_i}^\Pi = \beta_{B_jB_i}^\Pi + {\beta^\prime}_{B_jB_i}^\Pi$ is the contribution from the Feynman
graph in Fig.~\ref{fig:eins}(a), $\bar{\lambda}_{B_jB_i}^\Pi = \lambda_{B_jB_i}^\Pi + {\lambda^\prime}_{B_jB_i}^\Pi$ is the one-loop correction due to wavefunction
renormalization, Figs.~\ref{fig:eins}(b,c),
\begin{eqnarray}
\sqrt{Z_{B_j}Z_{B_i}} = 1 - \sum_\Pi \bar{\lambda}_{B_jB_i}^\Pi F^{(1)}(m_\Pi,0,\mu), \qquad \bar{\lambda}_{B_jB_i}^\Pi = \frac12 (\bar{\lambda}_{B_i}^\Pi +
\bar{\lambda}_{B_j}^\Pi),
\end{eqnarray}
and $\gamma_{B_jB_i}^\Pi$ is the correction arising from Fig.~\ref{fig:eins}(d). Here $\Pi$ stands for $\pi$, $K$, and $\eta$ mesons and $F^{(1)}(m_\Pi, 0, \mu)$ and
$I(m_\Pi, \mu)$ denote the loop functions defined in Eqs.~(\ref{eq:fprime}) and (\ref{eq:intd}). The unprimed and primed quantities are contributions with intermediate
octet and decuplet baryons, respectively. Finally, $u$ is a spinor referring to the initial and final baryon states $B_i$ and $B_j$. The explicit formulas for the chiral
coefficients $\alpha_{B_jB_i}$, $\bar{\beta}_{B_jB_i}^\Pi$, $\bar{\lambda}_{B_jB_i}^\Pi$, and $\gamma_{B_jB_i}^\Pi$ are listed in Appendix \ref{appB} for the sake of
completeness. Observe that if we restrict ourselves to the case of nonanalytic corrections in the limit $m_u=m_d=0$, and use the Gell-Mann--Okubo mass formula to rewrite
$m_\eta^2$ as $(4/3)m_K^2$, Eq.~(\ref{eq:axren}) reduces to results already obtained \cite{jm255,jm259}.

In close analogy to Eq.~(\ref{eq:pisym}), Eq.~(\ref{eq:axren}) can also be split into flavor singlet, flavor octet and flavor $\mathbf{27}$ contributions in terms of flavor
singlet, flavor octet, and flavor $\mathbf{27}$ linear combinations of $F^{(1)}(m_\Pi,0,\mu)$ and $I(m_\Pi,\mu)$. Thus, in order to keep our formulas compact, the renormalized
baryon axial vector current can be cast into the form
\begin{eqnarray}
\langle B_j | J_\mu^A | B_i \rangle & = & \left[\alpha_{B_jB_i} + b_\mathbf{1}^{B_jB_i} F_\mathbf{1}^{(1)} + b_\mathbf{8}^{B_jB_i}F_\mathbf{8}^{(1)} + b_\mathbf{27}^{B_jB_i}F_
\mathbf{27}^{(1)} + c_\mathbf{1}^{B_jB_i}I_\mathbf{1} + c_\mathbf{8}^{B_jB_i}I_\mathbf{8} + c_\mathbf{27}^{B_jB_i}I_\mathbf{27} \right] \nonumber \\
&  & \mbox{} \times \bar{u}_{B_j} \gamma_\mu \gamma_5 u_{B_i}, \label{eq:axv2}
\end{eqnarray}
where the new coefficients are
\begin{eqnarray}
b_\mathbf{1}^{B_jB_i} & = & -(a_{B_jB_i}^\pi + a_{B_jB_i}^K + a_{B_jB_i}^\eta), \label{eq:b01} \\
b_\mathbf{8}^{B_jB_i} & = & -\frac{1}{\sqrt 3} \left(a_{B_jB_i}^\pi - \frac12 a_{B_jB_i}^K - a_{B_jB_i}^\eta\right), \label{eq:b08} \\
b_\mathbf{27}^{B_jB_i} & = & -\frac{3}{40} \left(a_{B_jB_i}^\pi - 3a_{B_jB_i}^K + 9 a_{B_jB_i}^\eta\right), \label{eq:b27} \\
c_\mathbf{1}^{B_jB_i} & = & \gamma_{B_jB_i}^\pi + \gamma_{B_jB_i}^K + \gamma_{B_jB_i}^\eta, \label{eq:c01} \\
c_\mathbf{8}^{B_jB_i} & = & \frac{1}{\sqrt 3} \left(\gamma_{B_jB_i}^\pi - \frac12 \gamma_{B_jB_i}^K - \gamma_{B_jB_i}^\eta\right), \label{eq:c08} \\
c_\mathbf{27}^{B_jB_i} & = & \frac{3}{40} \left(\gamma_{B_jB_i}^\pi - 3\gamma_{B_jB_i}^{K} + 9 \gamma_{B_jB_i}^{\eta}\right), \label{eq:c27}
\end{eqnarray}
and the various $a_{B_jB_i}^\Pi$ are expressed in terms of the chiral coefficients as
\begin{eqnarray}
a_{B_jB_i}^\Pi = \bar{\beta}_{B_jB_i}^\Pi - \bar{\lambda}_{B_jB_i}^\Pi \alpha_{B_jB_i}.
\end{eqnarray}
Equations (\ref{eq:b01})-(\ref{eq:c27}) will be particularly useful in the comparison with the results obtained in the framework of large-$N_c$ heavy baryon chiral
perturbation theory. This will be done in the next section.

\section{Comparison between the two approaches in the limit $\Delta/m_\Pi =0$}
\label{comparison}

The matrix elements of the space components of the renormalized baryon axial vector current between initial and final baryon states $B_i$ and $B_j$ can be denoted as
\begin{equation}
\langle B_j|\bar{\psi}\gamma^k \gamma_5 T^c\psi|B_i \rangle = [A_\text{ren}^{kc}]_{B_jB_i}.
\end{equation}
Here $A_\text{ren}^{kc}=A^{kc}+\delta A^{kc}$, $\psi$ are the QCD quark fields, and $B_i$ and $B_j$ are baryons in the lowest-lying irreducible representation of
contracted $SU(6)$ spin-flavor symmetry, namely, the spin-$\frac12$ octet and the spin-$\frac32$ decuplet baryons. If the initial and final baryon states are restricted to
the spin-$\frac12$ octet baryons, the matrix elements $[A_\text{ren}^{kc}]_{B_jB_i}$ yield the actual values of $g_A^{B_jB_i}$, the axial vector couplings of the baryons.

In the degeneracy limit the renormalization to the baryon axial vector current reads
\begin{equation}
\delta A_{\text{deg}}^{kc} = \frac12 \left[A^{ja}, \left[A^{jb},A^{kc}
\right]\right] \Pi_{(1)}^{ab} - \frac12 \left[T^a, \left[T^b,A^{kc}
\right]\right] \Pi^{ab}. \label{eq:deg}
\end{equation}
At the physical value $N_c=3$, there is a one-to-one correspondence between the different flavor contributions of $[A_\text{ren}^{kc}]_{B_jB_i}$ and those contained
in Eq.~(\ref{eq:axv2}). The comparison can be made through
\begin{eqnarray}
&  & \left[ \frac12 [A^{ia}, [A^{ia},A^{kc}]] \right]_{B_jB_i} = b_\mathbf{1}^{B_jB_i}, \label{eq:cb1} \\
&  & \left[ \frac12  d^{ab8} [A^{ia}, [A^{ib},A^{kc}]] \right]_{B_jB_i} = b_\mathbf{8}^{B_jB_i}, \label{eq:cb8} \\
&  & \left[ \frac12 [A^{i8}, [A^{i8},A^{kc}]] \right]_{B_jB_i} = b_\mathbf{27}^{B_jB_i}, \label{eq:cb27} \\
&  & -\left[ \frac12 [T^a, [T^a,A^{kc}]] \right]_{B_jB_i} = c_\mathbf{1}^{B_jB_i}, \label{eq:cc1} \\
&  & -\left[ \frac12  d^{ab8} [T^a, [T^b,A^{kc}]] \right]_{B_jB_i} = c_\mathbf{8}^{B_jB_i}, \label{eq:cc8} \\
&  & -\left[ \frac12 [T^8, [T^8,A^{kc}]] \right]_{B_jB_i} = c_\mathbf{27}^{B_jB_i}. \label{eq:cc27}
\end{eqnarray}
It is understood that flavor singlet and octet pieces must be subtracted off Eqs.~(\ref{eq:cb27})~and~(\ref{eq:cc27}) in order to have a truly $\mathbf{27}$
contribution.

For instance, for the process $n \rightarrow p + e + \bar{\nu}_e$, the singlet component of the renormalized axial vector coupling -- diagrams~\ref{fig:eins}(a,b,c) --
reads (see Appendix \ref{app:mtx}),
\begin{eqnarray}
\left[ \frac12 [A^{ia}, [A^{ia},A^{kc}]] \right]_{pn} & = & \frac{115}{144} a_1^3 + \frac{7}{48} a_1^2b_2 +
\frac{19}{48} a_1b_2^2 - \frac{31}{432} a_1^2b_3 - \frac{11}{12} a_1^2 c_3 \nonumber \\
&  & \mbox{} + \frac{7}{144} b_2^2 + \frac{169}{216} a_1b_2b_3 - \frac{37}{36} a_1b_2c_3.
\end{eqnarray}
To the order of approximation implemented in this work, this corresponds exactly to $b_\mathbf{1}^{pn}$, Eq.~(\ref{eq:cb1}), given in terms of
$\alpha_{pn}$, $\bar{\beta}_{pn}$, and $\bar{\lambda}_{pn}$ , whose explicit expressions can be found in Appendix \ref{appB}. Note that, in order to make the comparison,
the baryon-meson couplings have to be expressed in terms of the coefficients of the $1/N_c$ expansion at $N_c=3$ as \cite{jen96}
\begin{eqnarray}
\begin{array}{l}
\displaystyle
D = \frac12 a_1 + \frac16 b_3, \\ [3mm]
\displaystyle
F = \frac13 a_1 + \frac16 b_2 + \frac19 b_3, \\ [3mm]
\displaystyle
\mathcal{C} = - a_1 - \frac12 c_3, \\ [3mm]
\displaystyle
\mathcal{H} = - \frac32 a_1 - \frac32 b_2 - \frac52 b_3.
\end{array} \label{eq:cid}
\end{eqnarray}
The agreement between the two approaches can be seen term by term in all expressions given by Eqs.~(\ref{eq:cb1}) to (\ref{eq:cc27}): Both approaches yield the same
results. An analogous comparison for the baryon mass relations, using the above identifications, was performed in Ref.~\cite{jen96}.

To close this section, a fit to baryon semileptonic decays by using the measured decay rates and
$g_A/g_V$ ratios \cite{part} is performed. Our motivation here is not really to be definitive about the
predictions of our expressions for $g_A$ but rather to explore the quality of our working assumptions. To the order
of approximation we implemented here, the fit \footnote{The quoted errors of the best fit parameters will be from the
$\chi^2$ fit only and will not include any theoretical uncertainties.} gives
$a_1=0.32\pm 0.04$, $b_2=-0.46\pm 0.03$, $b_3=3.04\pm 0.13$, and $c_3= 2.49$, with $\chi^2=38.18$ for 11 degrees of freedom,
or equivalently $F=0.37\pm 0.01$, $D=0.66 \pm 0.01$, and $\mathcal{H}=-7.39\pm 0.25$. The proton matrix element of the
$T^8$ component of the axial vector current (which is equal to $3F-D$ in the $SU(3)$ symmetry limit) is found
to be $0.45\pm 0.01$, which is smaller than its $SU(6)$ symmetric value of 1.
The coefficient $c_3$ was determined indirectly through
the relation $|\mathcal{C}| \sim 1.6$, which was obtained by a fit to the $\Delta\to N\pi$ decay rate \cite{jm259}.
It should be pointed out that the coupling $\mathcal{H}$ obtained in the fit is not close to its $SU(6)$ value, which is $3D-9F$;
this is mainly due to the order of approximation used here.

The predicted values of $g_A$ are listed in Table \ref{t:ga}, where the different flavor contributions are given separately.
As one might have anticipated, the $\mathbf{27}$ contribution to $g_A$ is suppressed relative to the octet contribution, which in turn is suppressed
relative to the singlet one. It is also instructive to remark that the highest contributions to $\chi^2$ come from the decay rate and
$g_A/g_V$ ratio of the process $\Xi^- \to \Lambda e^- \overline \nu_e$ (18.91 and 7.46, respectively), which
might suggest some inconsistencies in these data.

Evidently, a more complete analysis which can yield a better fit should also incorporate seven-body operators -- like $G\mathcal{D}_3\mathcal{D}_3$ -- or higher in the
correction to the baryon axial vector current (\ref{eq:deglim}). These terms represent $\mathcal{O}(1/{N_c^3})$ corrections or higher to the tree-level result
$\mathcal{O}(N_c)$. Although a substantial improvement of the value of $\mathcal{H}$, for instance, is expected,
the algebraic manipulations to reduce the double commutator
$[A^{ia},[A^{ib},A^{kc}]]\Pi_{(1)}^{ab}$ to the operator basis require a formidable effort which goes beyond the scope of the present paper.
One can also, of course, follow a more pragmatic approach and evaluate directly the matrix elements of the double commutator between octet
baryon states and observe the agreement with heavy baryon chiral perturbation theory pointed out above. This procedure, however,
does not allow to show the large-$N_c$ cancellations explicitly.

\begingroup
\begin{table}
\caption{\label{t:ga}Values of $g_A$ for various semileptonic processes.}
\begin{center}
\begin{tabular}{lccccc}
\hline\hline
Process & Total value & Tree level & Singlet piece & Octet piece & $\mathbf{27}$ piece \\ \hline
$n \to p e^- \overline \nu_e$             &   1.272  &   1.031  &   0.279  & $-0.040$ &   0.002 \\
$\Sigma^+ \to \Lambda e^+ \nu_e$          &   0.653  &   0.542  &   0.168  & $-0.057$ &   0.000 \\
$\Sigma^- \to \Lambda e^-\overline \nu_e$ &   0.624  &   0.542  &   0.113  & $-0.031$ & $-0.000$ \\
$\Lambda \to p e^- \overline \nu_e$       & $-0.904$ & $-0.720$ & $-0.134$ & $-0.055$ &   0.005 \\
$\Sigma^- \to n e^-\overline \nu_e$       &   0.375  &   0.298  &   0.080  & $-0.002$ & $-0.001$ \\
$\Xi^- \to \Lambda e^- \overline \nu_e$   &   0.139  &   0.178  & $-0.034$ & $-0.004$ & $-0.001$ \\
$\Xi^-\to \Sigma^0 e^- \overline \nu_e$   &   0.869  &   0.729  &   0.128  &   0.014  & $-0.002$ \\
$\Xi^0\to \Sigma^+ e^- \overline \nu_e$   &   1.312  &   1.031  &   0.246  &   0.041  & $-0.006$ \\
\hline\hline
\end{tabular}
\end{center}
\end{table}
\endgroup

\section{Inclusion of the $\bm{\eta^\prime}$}
\label{etaPrime}

So far, the renormalization of the baryon axial vector current has been performed by taking into account the contribution of the octet mesons in the loops,
Eq.~(\ref{eq:dakc}). In the large-$N_c$ limit, however, the quark loop responsible for the axial $U(1)$ anomaly is suppressed and the chiral symmetry is extended from
$SU(3)_R\times SU(3)_L\times U(1)_V$ to $U(3)_R\times U(3)_L$. As a consequence, the contribution from the $\eta^\prime$ should be included in the analysis.

Planar QCD flavor symmetry implies that the baryon $1/N_c$ chiral Lagrangian (\ref{eq:ncch}) possesses a $SU(2)\times U(3)$ spin-flavor symmetry at leading order in the
$1/N_c$ expansion and constrains this Lagrangian by forming a nonet baryon axial vector current out of the singlet and octet baryon axial vector currents at leading order
in the $1/N_c$ expansion \cite{jen96}, namely,
\begin{equation}
A^k = A^{k9} + \mathcal{O}(1/N_c), \label{eq:csin}
\end{equation}
where $A^k$ is the flavor singlet baryon axial vector current given in Eq.~(\ref{eq:asin}).
In Ref.~\cite{jen96} the constraint (\ref{eq:csin}) was imposed through the relation
\begin{equation}
b_n^{1,1} \rightarrow \overline{b}_n^{1,1} + \frac{1}{N_c} b_n^{1,1} \, ,
\end{equation}
where the coefficients $\overline{b}_n^{1,1}$ are determined by exact nonet symmetry, whereas the others are not constrained and violate nonet symmetry at first subleading
order $1/N_c$. Thus, for $N_c =3$, nonet symmetry implies that
\begin{subequations}
\begin{eqnarray}
\overline{b}_1^{1,1} & = & \frac{1}{\sqrt{6}}(a_1+b_2), \\
\overline{b}_3^{1,1} & = & \frac{1}{\sqrt{6}}(2b_3), 
\end{eqnarray}
\end{subequations}
where $a_1$, $b_2$, and $b_3$ are the operator coefficients of the octet axial vector current expansion Eq.~(\ref{eq:akc}). The above relations can be easily obtained by
using the ninth flavor components of $G^{ia}$ and $T^a$ given by \cite{jen96}
\begin{equation}
\label{T9}
G^{i9} = \frac{1}{\sqrt{6}} J^i, \qquad T^9 = \frac{1}{\sqrt{6}} N_c \openone.
\end{equation}
One should notice that the coefficients of the diagonal operators $\mathcal{D}_n^i$ in the singlet expansion do not depend
on the coefficients $c_n$ of the off-diagonal operators $\mathcal{O}_n^{ia}$ of the octet expansion.

The inclusion of the $\eta^\prime$ meson into the renormalization of $A^{kc}$ is now straightforwardly obtained in the degeneracy limit. Let us first discuss the
contribution from diagrams ~\ref{fig:eins}(a,b,c):
\begin{equation}
\delta A^{kc} = \frac12 [A^{i9},[A^{i9},A^{kc}]] F^{(1)} (m_{\eta^\prime},0,\mu). \label{eq:cnon}
\end{equation}
To the order of approximation implemented here, one has to evaluate the following commutator-anticommutator structures:
\begin{equation}
[J^i,[J^i,A^{kc}]] = 2 A^{kc},
\end{equation}
\begin{equation}
[J^i,[\{J^2,J^i\},G^{kc}]]+[\{J^2,J^i\},[J^i,G^{kc}]] =2 \mathcal{D}_3^{kc} + 8 \mathcal{O}_3^{kc},
\end{equation}
and
\begin{equation}
[J^i,[\{J^2,J^i\},\mathcal{D}_2^{kc}]]+[\{J^2,J^i\},[J^i,\mathcal{D}_2^{kc}]] =4 \mathcal{D}_4^{kc}.
\end{equation}
The correction due to the inclusion of the  $\eta^\prime$ thus amounts to
\begin{eqnarray}
\delta A^{kc} & = & \frac16 \left[ a_1 (\overline{b}_1^{1,1})^2G^{kc} + \frac{1}{N_c} b_2 (\overline{b}_1^{1,1})^2 \mathcal{D}_2^{kc} +
 \frac{1}{N_c^2} b_3 (\overline{b}_1^{1,1})^2 \mathcal{D}_3^{kc} + \frac{1}{N_c^2} c_3 (\overline{b}_1^{1,1})^2 \mathcal{O}_3^{kc} \right. \nonumber \\
&  & \mbox{} + \left. \frac{1}{N_c^2} a_1 (\overline{b}_1^{1,1})(\overline{b}_3^{1,1})\left(\mathcal{D}_3^{kc}+4\mathcal{O}_3^{kc} \right)
+ \frac{2}{N_c^3} b_2 (\overline{b}_1^{1,1}) (\overline{b}_3^{1,1})\mathcal{D}_4^{kc} \right] F^{(1)}(m_{\eta^\prime},0,\mu).
\end{eqnarray}

On the other hand, as far as conventional baryon chiral perturbation theory (i.e., without $1/N_c$-expansion) is concerned, the flavor singlet baryon-$\eta^\prime$ couplings
can be incorporated into the chiral effective Lagrangian Eq.~(\ref{eq:efflag}) by adding the two terms \cite{jen96}
\begin{equation}
2S_B \text{Tr} \mathcal{A}_\mu \text{Tr} \overline{B}_v S_v^\mu B_v -
2S_T \text{Tr}\mathcal{A}_\nu \overline{T}_v^\mu S_v^\nu T_{v \mu}, \label{eq:accc}
\end{equation}
where $S_B$ and $S_T$  are the singlet axial vector coupling constants of the octet and decuplet, respectively. The condition of nonet symmetry for the baryon axial
vector couplings implies
\begin{equation}
\label{eq:SbSt}
S_B \rightarrow \frac13 (3F-D), \qquad \qquad S_T \rightarrow -\frac13 {\cal H}.
\end{equation}
The contribution of the $\eta^\prime$ meson to the correction (\ref{eq:axren}) can be written as
\begin{equation}
\delta \langle B_j | J_\mu^A | B_i \rangle =  \left[\zeta_{B_jB_i}^{\eta^\prime}F^{(1)} (m_{\eta^\prime},0,\mu)\right]
\bar{u}_{B_j} \gamma_\mu \gamma_5 u_{B_i}, \label{eq:ccnon}
\end{equation}
where $\zeta_{B_jB_i}^{\eta^\prime}$ are the chiral coefficients which emerge from Fig.~\ref{fig:eins}(a,b,c) and can be found in Appendix \ref{appB}.

As in the previous section, a direct comparison between Eqs.~(\ref{eq:cnon}) and (\ref{eq:ccnon}) can be performed. In this case,
the comparison can be made through
\begin{equation}
\left[ \frac12 [A^{i9}, [A^{i9},A^{kc}]] \right]_{B_jB_i} = \, \zeta_{B_jB_i}^{\eta^\prime}, \label{eq:etappp}
\end{equation}
by using the identifications (\ref{eq:cid}) and (75). We have checked that, for the eight decays considered in the present study, the two approaches yield
the same result.

Finally, we briefly discuss the remaining diagram ~\ref{fig:eins}(d). The corresponding one-loop correction to the baryon axial vector current in large-$N_c$ chiral
perturbation theory was derived in Sec.~\ref{renormalization}, Eq.~(\ref{corr1d}). Including the $\eta^\prime$ thus amounts to the extra term
\begin{eqnarray}
\delta A^{kc} = - \frac12 \left[T^9, \left[T^9,A^{kc}\right]\right] \, I(m_{\eta^\prime},\mu).
\end{eqnarray}
However, the flavor operator $T^9$ is proportional to the unit matrix (\ref{T9}), such that the commutators are zero and there is thus no contribution from
diagram ~\ref{fig:eins}(d). Likewise, in conventional baryon chiral perturbation theory, the additional piece in the axial vector current due to the term involving $S_B$
in (\ref{eq:accc}) does not contribute. Again, in the degeneracy limit, the two approaches agree.

\section{Conclusions}
\label{conclusions}

In this paper we have computed the renormalization of the baryon axial vector current in the framework of heavy baryon chiral perturbation theory in the large-$N_c$ limit.
The analysis was performed at one-loop order, where the correction to the baryon axial vector current is given by an infinite series, each term representing a complicated
combination of commutators and/or anticommutators of the baryon axial vector current $A^{kc}$ and mass insertions $\mathcal{M}$. We have explicitly evaluated the first four
terms in this expansion: The contribution $AAA$ in the degeneracy limit $\Delta/m_{\Pi}=0$, the leading ($AAA\mathcal{M}$) and the two next-to-leading ($AAA\mathcal{MM}$)
order contributions for nonzero octet-decuplet mass difference, respectively. The general structure of these large-$N_c$ cancellations was already discussed in
Ref.~\cite{fmhjm}, where also a new large-$N_c$ cancellation in the singlet piece of the structure $AAA\mathcal{M}$ was identified.

Our motivation to go beyond this general analysis and to engage ourselves into the reduction of these rather involved operator products, including up to six $SU(6)$
spin-flavor operators $J^k, T^c$ and $G^{kc}$, was to explicitly demonstrate how these large-$N_c$ cancellations occur. It has already been pointed out in
Refs.~\cite{dm315,j315,djm94,fmhjm}, that there are large-$N_c$ cancellations between individual Feynman diagrams in the degeneracy limit, provided one sums over all
baryon states in a complete multiplet of the large-$N_c$ $SU(6)$ spin-flavor symmetry, i.e., over both the octet and decuplet, and uses axial coupling ratios given by the
large-$N_c$ spin-flavor symmetry. Indeed, our final expressions referring to the degeneracy limit explicitly demonstrate that the double commutator $AAA$ is of order
$N_c$ rather than of order $N_c^3$, as one would naively expect. As for the non-degenerate case we have shown that the new large-$N_c$ cancellation found in
Ref.~\cite{fmhjm} is a generic feature of the corresponding commutator-anticommutator structure $GGGJ^2$: The new cancellation observed in the singlet piece of $GGGJ^2$
indeed repeats itself in the octet and the $\mathbf{27}$. On the other hand, in the structure $GG\mathcal{D}_2J^2$, no new large-$N_c$ cancellations are detected: the
expression is of order $N_c^3$, consistent with the global analysis of Ref.~\cite{fmhjm}. However, in one of the two commutator-anticommutator structures $GGGJ^2J^2$ with
two mass insertions, a new large-$N_c$ cancellation was identified: Although naively one would expect this structure to be of order $N_c^3$, our explicit calculation for
the singlet, octet and $\mathbf{27}$ piece shows that it is of order $N_c^2$.

In the degeneracy limit, we have also performed a comparison of the
renormalized baryon axial vector current, obtained within two different
schemes: Large-$N_c$ baryon chiral perturbation theory on the one hand,
and conventional heavy baryon chiral perturbation theory (including both
octet and decuplet baryons), where no $1/N_c$ expansion is involved, on
the other hand. Both approaches agree -- the large-$N_c$ cancellations
are guaranteed to occur as a consequence of the contracted spin-flavor
symmetry present in the limit $N_c\!\to\!\infty$. By keeping the large-
$N_c$ spin-flavor symmetry manifest, one thus avoids large numerical
cancellations between loop diagrams with intermediate octet states and
low-energy constants of the next-to-leading order effective Lagrangian,
containing the effects of decuplet states \cite{meissner}.

In the present paper, we have taken into account the octet-decuplet mass difference, but neglected the $SU(3)$ splittings of the octet and decuplet baryons. Moreover, the
comparison between large-$N_c$ baryon chiral perturbation theory and conventional heavy baryon chiral perturbation theory, was performed for the degeneracy limit only.
The extension to the nondegenerate case, as well as the incorporation of $SU(3)$ mass splittings is currently in progress.

\acknowledgments

The authors would like to express their gratitude to Elizabeth
Jenkins and Aneesh V.\ Manohar for interesting discussions and comments
on the manuscript, as well as for their warm hospitality extended at UCSD, where
this work was initiated. This work has been partially supported by CONACYT (Mexico), under project SEP-2004-C01-47604. C.P.H.\ also acknowledges support by
the {\it Fondo Ram\'on Alvarez Buylla de Aldana} of Colima University.

\appendix

\section{Reduction of baryon operators\label{app:bo}}

Here we present the most general expressions, up to the order of approximation implemented in this work, for the two
commutator-anticommutator structures involved in the analysis.
The computation was performed by keeping $N_f$ and $N_c$ arbitrary, although the physical values $N_f=3$ and $N_c=3$ are used in the evaluation
of $g_A$.

\subsection{Degeneracy limit $\Delta/m_{\pi}=0$}

The flavor singlet, octet and {\bf 27} contributions of the double commutator
\begin{eqnarray}
[A^{ia}, [A^{ib}, A^{kc}]] \nonumber
\end{eqnarray}
can be organized as follows:

\textit{1.\ Flavor singlet contribution}
\begin{equation}
[G^{ia}, [G^{ia}, G^{kc}]] = \frac{3N_f^2-4}{4N_f} G^{kc},
\end{equation}
\begin{eqnarray}
&  & [G^{ia}, [G^{ia}, \mathcal{D}_2^{kc}]] + [G^{ia}, [ \mathcal{D}_2^{ia},G^{kc}]] + [ \mathcal{D}_2^{ia},[G^{ia},G^{kc} ]] = - \frac{2}{N_f} (N_c + N_f) G^{kc}
\nonumber \\
&  & \mbox {} + \frac{9N_f^2+8N_f-4}{4N_f} \mathcal{D}_2^{kc},
\end{eqnarray}
\begin{eqnarray}
&  & [G^{ia}, [\mathcal{D}_2^{ia}, \mathcal{D}_2^{kc}]] + [\mathcal{D}_2^{ia}, [G^{ia},\mathcal{D}_2^{kc}]] +
[\mathcal{D}_2^{ia}, [\mathcal{D}_2^{ia}, G^{kc}]] = \frac{N_c(N_c+2N_f)(N_f-2)-6N_f^2}{2N_f}G^{kc} \nonumber \\
&  & \mbox{} + \frac{2}{N_f}(N_c+N_f)(N_f-1)\mathcal{D}_2^{kc}
+ \frac{3N_f+2}{4} {\mathcal D}_3^{kc} + \frac{N_f}{2}{\mathcal O}_3^{kc},
\end{eqnarray}
\begin{eqnarray}
&  & [G^{ia}, [G^{ia}, {\mathcal D}_3^{kc}]] + [G^{ia}, [{\mathcal D}_3^{ia}, G^{kc}]] + [{\mathcal D}_3^{ia}, [G^{ia}, G^{kc}]] = [-N_c(N_c+2N_f)+2N_f-8] G^{kc}
\nonumber \\
&  & \mbox {} -3(N_c+N_f)\mathcal{D}_2^{kc} + \frac{13N_f^2+16N_f-12}{4N_f} {\mathcal D}_3^{kc} + \frac{N_f^2+2N_f-8}{N_f} {\mathcal O}_3^{kc},
\end{eqnarray}
\begin{eqnarray}
&  & [G^{ia}, [G^{ia}, {\mathcal O}_3^{kc}]] + [G^{ia}, [{\mathcal O}_3^{ia}, G^{kc}]] + [{\mathcal O}_3^{ia}, [G^{ia}, G^{kc}]] = [-N_c(N_c+2N_f)+N_f] G^{kc} \nonumber \\
&  & \mbox {} - \frac12(N_c+N_f) \mathcal{D}_2^{kc} + \frac{N_f+1}{2} {\mathcal D}_3^{kc} + \frac{15N_f^2+12N_f-4}{4N_f} {\mathcal O}_3^{kc},
\end{eqnarray}
\begin{eqnarray}
&  & [\mathcal{D}_2^{ia},[\mathcal{D}_2^{ia},\mathcal{D}_2^{kc}]]= \frac{N_c(N_c+2N_f)(N_f-2)-2N_f^2}{2N_f}\mathcal{D}_2^{kc}+\frac{N_f+2}{2}\mathcal{D}_4^{kc},
\end{eqnarray}
\begin{eqnarray}
&  & [G^{ia},[\mathcal{D}_2^{ia},\mathcal{D}_3^{kc}]]+[G^{ia},[\mathcal{D}_3^{ia},\mathcal{D}_2^{kc}]]+[\mathcal{D}_2^{ia},[G^{ia},\mathcal{D}_3^{kc}]] +
[\mathcal{D}_2^{ia},[\mathcal{D}_3^{ia},G^{kc}]]+ [\mathcal{D}_3^{ia},[G^{ia},\mathcal{D}_2^{kc}]]  \nonumber \\
&  & \mbox{} + [\mathcal{D}_3^{ia},[\mathcal{D}_2^{ia},G^{kc}]] = -12 (N_c+N_f)G^{kc} + [N_c(N_c+2N_f)-2N_f+8]\mathcal{D}_2^{kc}  \nonumber \\
&  & \mbox{} + \frac{7N_f-4}{N_f}(N_c+N_f) \mathcal{D}_3^{kc} + \frac{2(3N_f-4)}{N_f}(N_c+N_f) \mathcal{O}_3^{kc} + \frac{3N_f^2-4N_f-4}{N_f}\mathcal{D}_4^{kc},
\end{eqnarray}
\begin{eqnarray}
&  & [G^{ia},[\mathcal{D}_2^{ia},\mathcal{O}_3^{kc}]]+[G^{ia},[\mathcal{O}_3^{ia},\mathcal{D}_2^{kc}]]+[\mathcal{D}_2^{ia},[G^{ia},\mathcal{O}_3^{kc}]] +
[\mathcal{D}_2^{ia},[\mathcal{O}_3^{ia},G^{kc}]]+ [\mathcal{O}_3^{ia},[G^{ia},\mathcal{D}_2^{kc}]]  \nonumber \\
&  & \mbox{} + [\mathcal{O}_3^{ia},[\mathcal{D}_2^{ia},G^{kc}]] = -\frac32 [N_c(N_c+2N_f)-8N_f]\mathcal{D}_2^{kc}-\frac92(N_c+N_f) \mathcal{D}_3^{kc}- \frac{2}{N_f} (N_c+N_f)
\mathcal{O}_3^{kc} \nonumber \\
&  & \mbox{} + (3N_f+10) \mathcal{D}_4^{kc}.
\end{eqnarray}

\textit{2.\ Flavor octet contribution}
\begin{equation}
d^{ab8} [G^{ia}, [G^{ib}, G^{kc}]] = \frac{3N_f^2-16}{8N_f}d^{c8e} G^{ke} + \frac{N_f^2-4}{2N_f^2} \delta^{c8} J^k,
\end{equation}
\begin{eqnarray}
&  & d^{ab8} \left([G^{ia}, [G^{ib},\mathcal{D}_2^{kc}]] + [G^{ia},[ \mathcal{D}_2^{ib},G^{kc}]] + [ \mathcal{D}_2^{ia},[G^{ib},G^{kc}]]\right) = - \frac{2}{N_f}(N_c+N_f)
d^{c8e} G^{ke} \nonumber \\
&  & \mbox{} + \frac{5N_f+8}{8}d^{c8e} \mathcal{D}_2^{ke} - \frac{2}{N_f} \{G^{kc},T^8\} + \frac{N_f^2+2N_f-4}{2N_f} \{G^{k8},T^c\} + \frac{N_f+2}{4} [J^2, [T^8,G^{kc}]]
\nonumber \\
&  & \mbox{} + \frac{(N_c+N_f)(N_f-2)}{N_f^2} \delta^{c8} J^k,
\end{eqnarray}
\begin{eqnarray}
&  & d^{ab8} \left([G^{ia},[\mathcal{D}_2^{ib},\mathcal{D}_2^{kc}]] + [\mathcal{D}_2^{ia},[G^{ib},\mathcal{D}_2^{kc}]] + [\mathcal{D}_2^{ia},[\mathcal{D}_2^{ib},G^{kc}]]
\right) = - \frac32 N_f d^{c8e} G^{ke} \nonumber \\
&  & \mbox{} + \frac{(N_c+N_f)(N_f-2)}{N_f} \{G^{k8},T^c\} + \frac{(N_c+N_f)(N_f-4)}{2N_f}\{G^{kc},T^8\} + \frac38 N_f d^{c8e}{\mathcal D}_3^{ke} \nonumber \\
&  & \mbox{} + \frac{N_f-2}{4} d^{c8e} {\mathcal O}_3^{ke} + \frac12 \{G^{kc},\{J^r,G^{r8}\}\} + \frac12 \{G^{k8},\{J^r,G^{rc}\}\} + \frac{N_f-2}{2N_f} \{J^k,\{T^c,T^8\}\}
\nonumber \\
&  & \mbox{} + \frac14 (N_c+N_f)[J^2,[T^8,G^{kc}]],
\end{eqnarray}
\begin{eqnarray}
&  & d^{ab8} \left([G^{ia},[G^{ib},{\mathcal D}_3^{kc}]] + [G^{ia},[{\mathcal D}_3^{ib},G^{kc}]] + [{\mathcal D}_3^{ia},[G^{ib},G^{kc}]]\right) = (N_f-8) d^{c8e}G^{ke}
\nonumber \\
&  & \mbox {} - \frac32(N_c+N_f) d^{c8e}\mathcal{D}_2^{ke} - (N_c+N_f)\{G^{kc},T^8\}+ \frac{5N_f^2+12N_f-16}{8N_f} d^{c8e} {\mathcal D}_3^{ke} \nonumber \\
&  & \mbox{} + \frac{N_f^2+2N_f-24}{2N_f} d^{c8e} {\mathcal O}_3^{ke} - \frac34\{J^k,\{T^c,T^8\}\} + (N_f+1) \{J^k,\{G^{rc},G^{r8}\}\} \nonumber \\
&  & \mbox{} + \frac{N_f-4}{N_f}\{G^{kc},\{J^r,G^{r8}\}\} + \frac{N_f^2+3N_f-4}{N_f}\{G^{k8},\{J^r, G^{rc}\}\}
+ \frac32 (N_c+N_f) [J^2, [T^8, G^{kc}]] \nonumber \\
&  & \mbox{} - \frac{3N_c(N_c+2N_f)-8N_f+16}{2N_f} \delta^{c8}J^k + \frac{N_f^2+3N_f-4}{N_f^2} \delta^{c8} \{J^2,J^k\},
\end{eqnarray}
\begin{eqnarray}
&  & d^{ab8} \left([G^{ia},[G^{ib},{\mathcal O}_3^{kc}]] + [G^{ia},[{\mathcal O}_3^{ib},G^{kc}]] + [{\mathcal O}_3^{ia},[G^{ib},G^{kc}]] \right) = \frac{N_f}{2}
d^{c8e}G^{ke} \nonumber \\
&  & \mbox{} - \frac14 (N_c+N_f) d^{c8e} \mathcal{D}_2^{ke} - (N_c+N_f)\{G^{kc},T^8\} + \frac{N_f^2+N_f-8}{4N_f} d^{c8e} {\mathcal D}_3^{ke} \nonumber \\
&  & \mbox{} + \frac{7N_f^2+8N_f-16}{8N_f} d^{c8e} {\mathcal O}_3^{ke} - \frac18 \{J^k,\{T^c,T^8\}\} - \frac{N_f^2+N_f-8}{2N_f} \{J^k,\{G^{rc},G^{r8}\}\} \nonumber \\
&  & \mbox{} + (N_f + 2) \{G^{kc},\{J^r,G^{r8}\}\} - \frac{N_f+2}{2} \{G^{k8},\{J^r,G^{rc}\}\} - \frac{N_c(N_c+2N_f)}{4N_f} \delta^{c8} J^k \nonumber \\
&  & \mbox{} + \frac{2N_f^2+N_f-8}{2N_f^2} \delta^{c8}\{J^2,J^k\} - (N_c+N_f) [J^2, [T^8, G^{kc}]],
\end{eqnarray}
\begin{eqnarray}
&  & d^{ab8} [\mathcal{D}_2^{ia},[\mathcal{D}_2^{ib},\mathcal{D}_2^{kc}]] = - \frac{N_f}{2}d^{c8e}\mathcal{D}_2^{ke} + \frac{N_f-4}{4N_f}(N_c+N_f) \{J^k,\{T^c,T^8\}\}
+ \frac{N_f}{4} d^{c8e}\mathcal{D}_4^{ke} \nonumber \\
&  & \mbox{} + \{\mathcal{D}_2^{kc},\{J^r,G^{r8}\}\},
\end{eqnarray}
\begin{eqnarray}
&  & d^{ab8} \left([G^{ia},[ \mathcal{D}_2^{ib},\mathcal{D}_3^{kc}]]+[G^{ia},[\mathcal{D}_3^{ib},\mathcal{D}_2^{kc}]]+[\mathcal{D}_2^{ia},[G^{ib},\mathcal{D}_3^{kc}]] +
[\mathcal{D}_2^{ia},[\mathcal{D}_3^{ib},G^{kc}]] \right. \nonumber \\
&  & \mbox{} + \left. [\mathcal{D}_3^{ia},[G^{ib},\mathcal{D}_2^{kc}]] + [\mathcal{D}_3^{ia},[\mathcal{D}_2^{ib},G^{kc}]] \right) = - 6(N_c+N_f)d^{c8e}G^{ke} - (3N_f-6)
d^{c8e}\mathcal{D}_2^{ke} \nonumber \\
&  & \mbox{} + 2(N_f+1)\{G^{k8},T^c\} - 6\{G^{kc},T^8\} + \frac{N_f^2+8}{N_f} [J^2,[T^8,G^{kc}]]
+ \frac32(N_c+N_f)d^{c8e}\mathcal{D}_3^{ke} \nonumber \\
&  & \mbox{} + \frac{N_f-4}{N_f}(N_c+N_f)d^{c8e}\mathcal{O}_3^{ke} + \frac32(N_f-2)d^{c8e}\mathcal{D}_4^{ke}
+ \frac{N_c+N_f}{2} \{J^k,\{T^c,T^8\}\} \nonumber \\
&  & \mbox{} + \frac{2}{N_f}(N_f-2)(N_c+N_f)\{G^{kc},\{J^r,G^{r8}\}\}+ \frac{2}{N_f}(N_f-2)(N_c+N_f)\{G^{k8},\{J^r,G^{rc}\}\} \nonumber \\
&  & \mbox{} + \frac{N_f^2+3N_f-8}{N_f}\{J^2,\{G^{k8},T^c\}\}+\frac{3N_f-8}{N_f} \{J^2,\{G^{kc},T^8\}\} + 4\{\mathcal{D}_2^{k8},\{J^r,G^{rc}\}\}\nonumber \\
&  & \mbox{} -(N_f+4) \{\mathcal{D}_2^{kc},\{J^r,G^{r8}\}\} + \frac{5}{64}\{J^2,[G^{k8},\{J^r,G^{rc}\}]\}
- \frac{5}{64}\{J^2,[G^{kc},\{J^r,G^{r8}\}]\}
\nonumber \\
&  & \mbox{} - \frac{5}{64}\{[J^2,G^{kc}],\{J^r,G^{r8}\}\} + \frac{5}{64}\{[J^2,G^{k8}],\{J^r,G^{rc}\}\}
+ \frac{5}{64}\{J^k,[\{J^m,G^{mc}\},\{J^r,G^{r8}\}]\}
\nonumber \\
&  & \mbox{} + \frac{N_f^2+4N_f-8}{2N_f}\{J^2,[J^2,[T^8,G^{kc}]]\},
\end{eqnarray}
\begin{eqnarray}
&  & d^{ab8} \left([G^{ia},[\mathcal{D}_2^{ib},\mathcal{O}_3^{kc}]]+[G^{ia},[\mathcal{O}_3^{ib},\mathcal{D}_2^{kc}]]+[\mathcal{D}_2^{ia},[G^{ib},\mathcal{O}_3^{kc}]] +
[ \mathcal{D}_2^{ia},[\mathcal{O}_3^{ib},G^{kc}]] \right. \nonumber \\
&  & \mbox{} \left. + [\mathcal{O}_3^{ia},[G^{ib},\mathcal{D}_2^{kc}]] + [\mathcal{O}_3^{ia},[\mathcal{D}_2^{ib},G^{kc}]] \right) =
6 N_f d^{c8e}\mathcal{D}_2^{ke} -\frac{5N_f+8}{4N_f}(N_c+N_f)d^{c8e}\mathcal{D}_3^{ke} \nonumber \\
&  & \mbox{}  -\frac{2}{N_f}(N_c+N_f)d^{c8e}\mathcal{O}_3^{ke} - \frac{2}{N_f}(N_f-2)(N_c+N_f)\{J^k,\{G^{rc},G^{r8}\}\} \nonumber \\
&  & \mbox{} -\frac34(N_c+N_f)\{J^k,\{T^c,T^8\}\} + \frac{2}{N_f^2}(N_f-2)(N_c+N_f)\delta^{c8}\{J^2,J^k\} + \frac{N_f+9}{2}d^{c8e}\mathcal{D}_4^{ke} \nonumber \\
&  & \mbox{} + \frac{N_f^2+2N_f-4}{2N_f}\{J^2,\{G^{k8},T^c\}\} -\frac{2}{N_f} \{J^2,\{G^{kc},T^8\}\} -\frac{9N_f-4}{2N_f} \{\mathcal{D}_2^{k8},\{J^r,G^{rc}\}\} \nonumber \\
&  & \mbox{} + \frac{N_f^2+9N_f+4}{2N_f} \{\mathcal{D}_2^{kc},\{J^r,G^{r8}\}\} - \frac{71}{128}\{[J^2,G^{kc}],\{J^r,G^{r8}\}\} + \frac{71}{128}\{[J^2,G^{k8}],\{J^r,G^{rc}\}\}
\nonumber \\
&  & \mbox{} +\frac{71}{128}\{J^k,[\{J^m,G^{mc}\},\{J^r,G^{r8}\}]\} + \frac{N_f+2}{4}\{J^2,[J^2,[T^8,G^{kc}]]\}
\nonumber \\
&  & \mbox{} - \frac{71}{128}\{J^2,[G^{kc},\{J^r,G^{r8}\}]\} + \frac{71}{128} \{J^2,[G^{k8},\{J^r,G^{rc}\}]\},
\end{eqnarray}

\textit{3.\ Flavor {\bf 27}  contribution}
\begin{equation}
[G^{i8}, [G^{i8}, G^{kc}]] = \frac14 \left(f^{c8e}f^{8eg} + 2 d^{c8e} d^{8eg} \right) G^{kg} + \frac{1}{N_f} \delta^{c8} G^{k8} + \frac{1}{2N_f} d^{c88} J^k,
\end{equation}
\begin{eqnarray}
&  & [G^{i8}, [G^{i8}, \mathcal{D}_2^{kc}]] + [G^{i8}, [\mathcal{D}_2^{i8}, G^{kc}]] + [\mathcal{D}_2^{i8},[G^{i8}, G^{kc}]] = \left[ \frac{1}{N_f}\delta^{88}\delta^{cg}
+ \frac74 f^{c8e}f^{8eg}\right]\mathcal{D}_2^{kg} \nonumber \\
&  & \mbox{} + \frac{2}{N_f} \delta^{c8}\mathcal{D}_2^{k8} + d^{c8e} \{G^{ke}, T^8\} + \frac12 d^{88e} \{G^{ke}, T^c\} + if^{c8e} [G^{k8}, \{J^r, G^{re} \}] ,
\end{eqnarray}
\begin{eqnarray}
&  & [G^{i8}, [\mathcal{D}_2^{i8},\mathcal{D}_2^{kc}]] + [\mathcal{D}_2^{i8}, [G^{i8},\mathcal{D}_2^{kc}]] + [\mathcal{D}_2^{i8}, [\mathcal{D}_2^{i8}, G^{kc}]] =
-2 f^{c8e} f^{8eg} G^{kg} \nonumber \\
&  & \mbox{} + \frac34 f^{c8e} f^{8eg} {\mathcal D}_3^{kg} + \frac12 f^{c8e} f^{8eg} {\mathcal O}_3^{kg} + \frac12  \{G^{kc}, \{T^8,T^8\}\} + \{G^{k8},\{T^c,T^8\}\}
\nonumber \\
&  & \mbox{} - \frac12 f^{c8e} \epsilon^{kim} \{T^e, \{J^i, G^{m8} \}\},
\end{eqnarray}
\begin{eqnarray}
&  & [G^{i8}, [G^{i8}, {\mathcal D}_3^{kc}]] + [G^{i8}, [{\mathcal D}_3^{i8}, G^{kc}]] + [{\mathcal D}_3^{i8}, [G^{i8}, G^{kc}]] = (d^{c8e} d^{8eg}
- 2f^{c8e} f^{8eg}) G^{kg}\nonumber \\
&  & \mbox{} + \frac{2}{N_f} \delta^{c8} G^{k8} + \frac{1}{N_f} d^{c88} J^k + \frac{2}{N_f} \delta^{88}{\mathcal D}_3^{kc}
+ \frac14 (3 f^{c8e} f^{8eg} + 2 d^{c8e} d^{8eg}) {\mathcal D}_3^{kg} + \frac{1}{N_f} \delta^{c8} \mathcal{D}_3^{k8} \nonumber \\
&  & \mbox{}+ d^{c8e} d^{8eg} {\mathcal O}_3^{kg}  - 2 \{G^{kc}, \{G^{r8}, G^{r8} \}\} + 2 \{G^{rc}, \{G^{r8}, G^{k8} \}\} + 4 d^{c8e} \{ G^{ke}, \{J^r, G^{r8} \}\}
\nonumber \\
&  & \mbox{} - d^{c8e} \{G^{k8}, \{J^r, G^{re} \}\} + d^{88e} \{G^{ke},\{J^r, G^{rc} \}\} - 3 d^{c8e} \{J^k, \{G^{re},G^{r8} \}\} \nonumber \\
&  & \mbox{} + d^{88e} \{J^k, \{G^{rc}, G^{re}\}\} + \frac{1}{N_f} d^{c88} \{J^2, J^k\} - \frac12 f^{c8e} \epsilon^{kim} \{T^e, \{J^i, G^{m8} \}\},
\end{eqnarray}
\begin{eqnarray}
&  & [G^{i8}, [G^{i8}, {\mathcal O}_3^{kc}]] + [G^{i8}, [{\mathcal O}_3^{i8}, G^{kc}]] + [{\mathcal O}_3^{i8}, [G^{i8}, G^{kc}]] = - \frac12 (d^{c8e} d^{8eg}
- 2 f^{c8e} f^{8eg}) G^{kg}\nonumber \\
&  & \mbox{} - \frac{1}{N_f} \delta^{c8} G^{k8} - \frac{1}{2N_f} d^{c88} J^k + \frac12 d^{c8e} d^{8eg} {\mathcal D}_3^{kg} + \frac{1}{N_f} \delta^{c8}\mathcal{D}_3^{k8}
+ \frac{2}{N_f} \delta^{88}{\mathcal O}_3^{kc} + \frac{5}{N_f} \delta^{c8} \mathcal{O}_3^{k8} \nonumber \\
&  & \mbox{} + \frac14(3f^{c8e} f^{8eg} + 4d^{c8e} d^{8eg})
{\mathcal O}_3^{kg} - \{G^{kc},\{G^{r8}, G^{r8}\}\} - \{G^{rc}, \{G^{r8}, G^{k8} \}\} \nonumber \\
&  & \mbox{} - d^{c8e} \{G^{ke},\{J^r, G^{r8} \}\} + \frac32 d^{c8e} \{G^{k8}, \{J^r, G^{re} \}\} - \frac12 d^{88e} \{G^{ke}, \{J^r, G^{rc} \}\} \nonumber \\
&  & \mbox{} + d^{88e} \{G^{kc},\{J^r, G^{re} \}\} - \frac12 d^{c8e} \{J^k, \{G^{re},G^{r8} \}\} - \frac12 d^{88e} \{J^k, \{G^{rc}, G^{re}\}\} \nonumber \\
&  & \mbox{}  + \frac{1}{N_f} d^{c88} \{J^2, J^k\} + \frac34 f^{c8e}
\epsilon^{kim} \{T^e, \{J^i, G^{m8}\}\},
\end{eqnarray}
\begin{eqnarray}
&  & [\mathcal{D}_2^{i8},[\mathcal{D}_2^{i8},\mathcal{D}_2^{kc}]] = -f^{c8e}f^{8eg}\mathcal{D}_2^{kg} + \frac12 f^{c8e}f^{8eg} \mathcal{D}_4^{kg}
+ \frac12 \{\mathcal{D}_2^{kc},\{T^8,T^8\}\},
\end{eqnarray}
\begin{eqnarray}
&  & [G^{i8},[\mathcal{D}_2^{i8},\mathcal{D}_3^{kc}]]+[G^{i8},[\mathcal{D}_3^{i8},\mathcal{D}_2^{kc}]]+[\mathcal{D}_2^{i8},[G^{i8},\mathcal{D}_3^{kc}]]+
[\mathcal{D}_2^{i8},[\mathcal{D}_3^{i8},G^{kc}]] \nonumber \\
&  & \mbox{} + [\mathcal{D}_3^{i8},[G^{i8},\mathcal{D}_2^{kc}]] + [\mathcal{D}_3^{i8},[\mathcal{D}_2^{i8},G^{kc}]] =
4if^{c8e}[G^{k8},\{J^r,G^{re}\}] - 4if^{c8e} [G^{ke},\{J^r,G^{r8}\}] \nonumber \\
&  & \mbox{} + 2d^{c8e} \{J^2,\{G^{ke},T^8\}\} + d^{88e}\{J^2,\{G^{ke},T^c\}\} - 2 d^{c8e} \{\mathcal{D}_2^{k8},\{J^r,G^{re}\}\} \nonumber \\
&  & \mbox{} - d^{88e} \{\mathcal{D}_2^{kc},\{J^r,G^{re}\}\} + 2\{\{J^r,G^{rc}\},\{G^{k8},T^8\}\} + 2\{\{J^r,G^{r8}\},\{G^{kc},T^8\}\}
\nonumber \\
&  & \mbox{} + 2 \{\{J^r,G^{r8}\},\{G^{k8},T^c\}\} + 2if^{c8e}\{J^2,[G^{ke},\{J^r,G^{r8}\}]\} - 2if^{c8e}\{\{J^r,G^{re}\},[J^2,G^{k8}]\}
\nonumber \\
&  & \mbox{} + 2if^{c8e}\{J^k,[\{J^i,G^{ie}\},\{J^r,G^{r8}\}]\},
\end{eqnarray}
\begin{eqnarray}
&  & [G^{i8},[\mathcal{D}_2^{i8},\mathcal{O}_3^{kc}]]+[G^{i8},[\mathcal{O}_3^{i8},\mathcal{D}_2^{kc}]]+[\mathcal{D}_2^{i8},[G^{i8},\mathcal{O}_3^{kc}]]+
[\mathcal{D}_2^{i8},[\mathcal{O}_3^{i8},G^{kc}]]  \nonumber \\
&  & \mbox{} + [\mathcal{O}_3^{i8},[G^{i8},\mathcal{D}_2^{kc}]] + [\mathcal{O}_3^{i8},[\mathcal{D}_2^{i8},G^{kc}]] = 9f^{c8e}f^{8eg} \mathcal{D}_2^{kg}
+ \frac{2}{N_f}\delta^{88}\mathcal{D}_4^{kc} + \frac92 f^{c8e}f^{8eg} \mathcal{D}_4^{kg} \nonumber \\
&  & \mbox{} + \frac{4}{N_f} \delta^{c8}\mathcal{D}_4^{k8} - 2\{\mathcal{D}_2^{kc},\{G^{r8},G^{r8}\}\} - 4\{\mathcal{D}_2^{k8},\{G^{rc},G^{r8}\}\}
+ d^{c8e}\{\mathcal{D}_2^{k8},\{J^r,G^{re}\}\} \nonumber \\
&  & \mbox{} + \frac12 d^{88e}\{\mathcal{D}_2^{kc},\{J^r,G^{re}\}\} + d^{c8e}\{J^2,\{G^{ke},T^8\}\} + \frac12 d^{88e}\{J^2,\{G^{ke},T^c\}\} \nonumber \\
&  & \mbox{} + 2if^{c8e} \{J^2,[G^{k8},\{J^r,G^{re}\}]\} - if^{c8e} \{J^2,[G^{ke},\{J^r,G^{r8}\}]\}
- if^{c8e} \{\{J^r,G^{r8}\},[J^2,G^{ke}]\} \nonumber \\
&  & \mbox{} + if^{c8e} \{\{J^r,G^{re}\},[J^2,G^{k8}]\} - 2if^{c8e}\{J^k,[\{J^i,G^{ie}\},\{J^r,G^{r8}\}]\},
\end{eqnarray}

\subsection{Non-degenerate case $\Delta/m_{\pi} \neq 0$}

Similarly, the evaluation of the commutator-anticommutator structure
\begin{eqnarray}
\left\{ A^{ja}, \left[A^{kc}, \left[ \mathcal{M}, A^{jb} \right] \right] \right\},\nonumber
\end{eqnarray}
which represents the leading contribution to the renormalized baryon axial vector current for finite octet-decuplet mass difference, yields the following terms:

\textit{1.\ Flavor singlet contribution}
\begin{equation}
\label{GJ2GG-singlet}
\{G^{ia}, [G^{kc},[J^2,G^{ia}]] \} = -\frac12 (N_f-2) G^{kc} + \frac12 (N_c+N_f)\mathcal{D}_2^{kc} - \frac12 \mathcal{D}_3^{kc} - \mathcal{O}_3^{kc},
\end{equation}
\begin{eqnarray}
\label{GJ2GD2-singlet}
&   & \{G^{ia}, [\mathcal{D}_2^{kc}, [J^2,G^{ia}]] \} + \{G^{ia}, [G^{kc},[J^2,\mathcal{D}_2^{ia}]] \} + \{\mathcal{D}_2^{ia}, [G^{kc},[J^2,G^{ia}]] \} =
2(N_c+N_f) G^{kc} \nonumber\\
&   & \mbox{} + \frac12 [N_c(N_c+2N_f)-9N_f-2]\mathcal{D}_2^{kc} + \frac12 (N_c+N_f) \mathcal{D}_3^{kc} - 2 \mathcal{D}_4^{kc},
\end{eqnarray}

\textit{2.\ Flavor octet contribution}
\begin{eqnarray}
\label{GJ2GG-octet}
&   & d^{ab8} \{G^{ia},[G^{kc},[J^2,G^{ib}]] \} = -\frac14 (N_f-4) d^{c8e} G^{ke} + \frac14 (N_c+N_f) d^{c8e}\mathcal{D}_2^{ke} - \frac14 d^{c8e} \mathcal{D}_3^{ke}
\nonumber \\
&   & \mbox{} - \frac12 d^{c8e} \mathcal{O}_3^{ke} - \frac12 \{G^{kc},\{J^r,G^{r8}\}\} + \frac{1}{N_f} \{G^{k8},\{J^r,G^{rc}\}\} + \frac18
\{J^k,\{T^c,T^8\}\} \nonumber \\
&   & \mbox{} - \frac{1}{N_f} \{J^k, \{G^{rc},G^{r8}\}\} + \frac14 (N_c+N_f) [J^2,[T^8,G^{kc}]] + \frac{N_c(N_c+2N_f)-2N_f+4}{4N_f}
\delta^{c8}J^k \nonumber \\
&   & \mbox{} - \frac{1}{2 N_f}\delta^{c8} \{J^2,J^k\},
\end{eqnarray}
\begin{eqnarray}
\label{GJ2GD2-octet}
&   & d^{ab8} \left(\{G^{ia}, [\mathcal{D}_2^{kc}, [J^2,G^{ib}]] \} + \{G^{ia},[G^{kc},[J^2, \mathcal{D}_2^{ib}]] \} + \{\mathcal{D}_2^{ia}, [G^{kc}, [J^2,G^{ib}]] \}
\right) = \nonumber \\
&   & \mbox{} (N_c+N_f) d^{c8e} G^{ke} - \frac{7N_f+4}{4} d^{c8e}\mathcal{D}_2^{ke} - \frac{N_f}{2} \{T^c,G^{k8}\} + \{G^{kc},T^8\} + \frac14 (N_c+N_f) d^{c8e} \mathcal{D}_3^{ke} \nonumber \\
&   & \mbox{} + \frac{N_f-2}{2N_f} (N_c+N_f) \left( \{J^k,\{G^{rc},G^{r8}\}\} -
\{G^{k8},\{J^r,G^{rc}\}\} \right) + \frac14 (N_c+N_f) \{J^k,\{T^c,T^8\}\} \nonumber \\
&   & \mbox{} - \frac{N_f^2+4}{4N_f} [J^2, [T^8,G^{kc}]] + \frac12 \{\mathcal{D}_2^{k8}, \{J^r,G^{rc}\}\} - \frac{N_f-2}{2N_f} \{J^2, \{G^{k8},T^c\}\} - \frac12 d^{c8e} \mathcal{D}_4^{ke} \nonumber \\
&   & \mbox{} - \frac{N_f+1}{N_f} \{\mathcal{D}_2^{kc}, \{J^r, G^{r8}\}\},
\end{eqnarray}

\textit{3.\ Flavor {\bf 27} contribution}
\begin{eqnarray}
\label{GJ2GG-27}
&   & \{G^{i8}, [G^{kc}, [J^2,G^{i8}]] \} = \frac{1}{N_f} \left(\delta^{88}\delta^{bc} - \delta^{b8} \delta^{c8} \right) G^{kb} - \frac12 \left(
d^{c8e} d^{e8d} - d^{ced} d^{e88} - f^{c8e} f^{e8d} \right) G^{kd} \nonumber \\
&   & \mbox{} - \frac12 d^{c8e} \{G^{k8},\{J^r,G^{re}\}\} + \frac12 d^{c8e} \{J^k,\{G^{re},G^{r8}\}\} - \frac14 \epsilon^{kij}
f^{c8e} \{T^e, \{J^i, G^{j8}\}\}  \nonumber \\
&   & \mbox{} + \frac{1}{2 N_f} \delta^{c8}\{J^k,\{J^r,G^{r8}\}\} - \frac{1}{N_f} \delta^{c8} \{J^2,G^{k8}\},
\end{eqnarray}
\begin{eqnarray}
\label{GJ2GD2-27}
&  & \{G^{i8}, [\mathcal{D}_2^{kc}, [J^2,G^{i8}]] \} + \{G^{i8}, [G^{kc},[J^2, \mathcal{D}_2^{i8}]] \} + \{\mathcal{D}_2^{i8}, [G^{kc}, [J^2,G^{i8}]] \} = \nonumber \\
&  & -\frac{15}{4}f^{c8e}f^{8eg}\mathcal{D}_2^{kg} + \frac{i}{2}f^{c8e}[G^{ke},\{J^r,G^{r8}\}] -
if^{c8e}[G^{k8},\{J^r,G^{re}\}] - \frac12 f^{c8e}f^{8eg}\mathcal{D}_4^{kg} \nonumber \\
&  & \mbox{} + \{\mathcal{D}_2^{kc},\{G^{r8},G^{r8}\}\} + \{\mathcal{D}_2^{k8},\{G^{rc},G^{r8}\}\}
- \frac12 \{\{J^r,G^{rc}\},\{G^{k8},T^8\}\} \nonumber \\
&  & \mbox{} - \frac12 \{\{J^r,G^{r8}\},\{G^{k8},T^c\}\} + \frac{i}{2}f^{c8e}\{J^k,[\{J^i,G^{ie}\},\{J^r,G^{r8}\}]\}.
\end{eqnarray}

\section{Matrix elements of baryon operators\label{app:mtx}}

In order to produce results of straightforward applicability, here we present the evaluation of the matrix elements of the baryon operators that constitute $A^{kc}$.
A glance at Eqs.~(\ref{eq:fsinglet})-(\ref{eq:f27}) reveals that one can identify the basic operators
\begin{eqnarray}
\begin{array}{lll}
X_0^c = \{J^r,G^{rc}\}, & X_1^{kc} = G^{kc}, & X_2^{kc} = \mathcal{D}_2^{kc}, \\
X_3^{kc}=\mathcal{D}_3^{kc}, & X_4^{kc} = \mathcal{O}_3^{kc}, & X_5^{kc} = \{G^{kc},T^8\}, \\
X_6^{kc} = \{G^{k8},T^c\}, & X_7^{kc} = \{G^{kc},\{J^r,G^{r8}\}\}, & X_8^{kc} = \{G^{k8},\{J^r,G^{rc}\}\}, \\
X_9^{kc} = \{J^k,\{G^{rc},G^{r8}\}\}, & X_{10}^{kc} = \{J^k,\{T^c,T^8\}\}, & X_{11}^{kc} = [G^{k8},\{J^r,G^{rc}\}], \\
X_{12}^{kc} = [G^{kc},\{J^r,G^{r8}\}], & X_{13}^{kc} = \{G^{kc},\{T^8,T^8\}\}, & X_{14}^{kc} = \{G^{k8},\{T^c,T^8\}\}, \\
X_{15}^{kc} = \{G^{rc},\{G^{r8},G^{k8}\}\}, & X_{16}^{kc} = \{G^{kc},\{G^{r8},G^{r8}\}\}, & X_{17}^{kc} = \{\mathcal{D}_2^{kc},\{G^{r8},G^{r8}\}\}, \\
X_{18}^{kc} = \{\mathcal{D}_2^{k8},\{G^{rc},G^{r8}\}\}. & &
\end{array} \nonumber
\end{eqnarray}

Among all the allowed operators, $\mathcal{O}_3^{kc}$ and $[J^2,G^{kc}]$ connect states of different spin only, whereas $[J^2,[T^8,G^{kc}]]$ connects states which change
both spin and strangeness and along with $f^{c8e} \epsilon^{kim}\{T^e,\{J^i,G^{m8}\}\}$, they do not contribute to any observed decay. Thus, the nonvanishing matrix elements
of the operators $X_m^{kc}$ for initial and final spin-$\frac12$ baryon states for eight physically relevant processes are listed in Table \ref{t:x1}. Notice that operators
of the form $f^{c8e}X_m^{ke}$, $d^{c8e}X_m^{ke}$, $f^{c8d}d^{d8e}X_m^{ke}, \dots$, can be trivially obtained from $X_m^{kc}$ and are not listed in Table \ref{t:x1}.

\begingroup
\begin{table}
\caption{\label{t:x1}Matrix elements of the operators $X_m^{kc}$ for some observed transitions between spin-$\frac12$ baryons.}
\begin{center}
\begin{tabular}{lcccccccc}
\hline\hline
& $pn$ & $\Lambda \Sigma^\pm$ & $\Xi^0\Xi^-$ & $p\Lambda$ & $n\Sigma^-$ & $\Lambda \Xi^-$ & $\Sigma^0\Xi^-$ & $\Sigma^+\Xi^0$ \\
\hline
$[X_0^c]_{B_jB_i}$ & $5/2$ & $\sqrt{3/2}$ & $1/2$ & $-\sqrt{27/8}$ & $1/2$ & $\sqrt{3/8}$ & $5\sqrt{2}/4$ & $5/2$ \\
\hline
$[X_1^{kc}]_{B_jB_i}$ & $5/6$ & $1/\sqrt{6}$ & $1/6$ & $-\sqrt{3/8}$ & $1/6$ & $1/\sqrt{24}$ & $5/\sqrt{72}$ & $5/6$ \\
$[X_2^{kc}]_{B_jB_i}$ & $1/2$ & $0$ & $-1/2$ & $-\sqrt{3/8}$ & $-1/2$ & $\sqrt{3/8}$ & $1/\sqrt{8}$ & $1/2$ \\
$[X_3^{kc}]_{B_jB_i}$ & $5/2$ & $\sqrt{3/2}$ & $1/2$ & $-\sqrt{27/8}$ & $1/2$ & $\sqrt{3/8}$ & $5/\sqrt{8}$ & $5/2$ \\
\hline
$[X_5^{kc}]_{B_jB_i}$ & $5/\sqrt{12}$ & $0$ & $-1/\sqrt{12}$ & $-3/\sqrt{32}$ & $1/\sqrt{48}$ & $-1/\sqrt{32}$ & $-5/\sqrt{96}$ & $-5/\sqrt{48}$ \\
$[X_6^{kc}]_{B_jB_i}$ & $1/\sqrt{12}$ & $0$ & $\sqrt{3}/2$ & $1/\sqrt{32}$ & $-\sqrt{3}/4$ & $-5/\sqrt{32}$ & $-1/\sqrt{96}$ & $-1/\sqrt{48}$ \\
$[X_7^{kc}]_{B_jB_i}$ & $5/\sqrt{48}$ & $0$ & $-\sqrt{3}/4$ & $3\sqrt{2}/16$ & $\sqrt{3}/8$ & $-5\sqrt{2}/16$ & $-5\sqrt{6}/48$ & $-5\sqrt{3}/24$ \\
$[X_8^{kc}]_{B_jB_i}$ & $5/\sqrt{48}$ & $0$ & $-\sqrt{3}/4$ & $3\sqrt{2}/16$ & $\sqrt{3}/8$ & $-5\sqrt{2}/16$ & $-5\sqrt{6}/48$ & $-5\sqrt{3}/24$ \\
$[X_9^{kc}]_{B_jB_i}$ & $5/\sqrt{48}$ & $-1/\sqrt{2}$ & $-11/\sqrt{48}$ & $3\sqrt{2}/16$ & $11\sqrt{3}/24$ & $-13\sqrt{2}/16$ & $-5\sqrt{6}/48$ & $-5\sqrt{3}/24$ \\
$[X_{10}^{kc}]_{B_jB_i}$ & $\sqrt{3}$ & $0$ & $\sqrt{3}$ & $-3/\sqrt{8}$ & $-\sqrt{3}/2$ & $-3/\sqrt{8}$ & $-\sqrt{3/8}$ & $-\sqrt{3}/2$ \\
\hline
$[X_{11}^{kc}]_{B_jB_i}$ & $0$ & $1/\sqrt{2}$ & $0$ & $-9\sqrt{2}/16$ & $-\sqrt{3}/24$ & $\sqrt{2}/16$ & $25\sqrt{6}/48$ & $25\sqrt{3}/24$ \\
$[X_{12}^{kc}]_{B_jB_i}$ & $0$ & $-1/\sqrt{2}$ & $0$ & $9\sqrt{2}/16$ & $\sqrt{3}/24$ & $-\sqrt{2}/16$ & $-25\sqrt{6}/48$ & $-25\sqrt{3}/24$ \\
$[X_{13}^{kc}]_{B_jB_i}$ & $5/2$ & $0$ & $1/2$ & $-\sqrt{27/32}$ & $1/4$ & $\sqrt{3/32}$ & $5/\sqrt{32}$ & $5/4$ \\
$[X_{14}^{kc}]_{B_jB_i}$ & $1/2$ & $0$ & $-3/2$ & $\sqrt{6}/16$ & $-3/8$ & $5\sqrt{6}/16$ & $\sqrt{2}/16$ & $1/8$ \\
$[X_{15}^{kc}]_{B_jB_i}$ & $5/72$ & $-\sqrt{6}/36$ & $97/72$ & $-5\sqrt{6}/96$ & $53/144$ & $101\sqrt{6}/288$ & $25\sqrt{2}/288$ & $25/144$ \\
$[X_{16}^{kc}]_{B_jB_i}$ & $5/24$ & $\sqrt{2/3}$ & $17/24$ & $-5\sqrt{6}/32$ & $13/48$ & $7\sqrt{6}/32$ & $145\sqrt{2}/96$ & $145/48$ \\
$[X_{17}^{kc}]_{B_jB_i}$ & $1/8$ & $0$ & $-17/8$ & $-5\sqrt{6}/32$ & $-13/16$ & $21\sqrt{6}/32$ & $29\sqrt{2}/32$ & $29/16$ \\
$[X_{18}^{kc}]_{B_jB_i}$ & $5/8$ & $0$ & $11/8 $ & $3\sqrt{6}/64$  & $11/32$  & $13\sqrt{6}/64$ & $5\sqrt{2}/64$ & $5/32$ \\
\hline\hline
\end{tabular}
\end{center}
\end{table}
\endgroup

We now proceed further to obtain theoretical expressions for the axial vector couplings $g_A^{B_jB_i}$. For any given process, $g_A^{B_jB_i}$ is composed of three terms.
The first one is the tree-level value $\alpha_{B_jB_i}$; the next one is the contribution of Figs.~\ref{fig:eins}(a,b,c); and the last one is the contribution of
Fig.~\ref{fig:eins}(d). The tree-level value can be written as a sum of the three parameters $a_1$, $b_2$, and $b_3$ times coefficients obtained from the appropriate
matrix elements of the baryon operators that accompany them; these coefficients are listed in Table \ref{t:x2} for the processes of interest here. The contribution of
Fig.~\ref{fig:eins}(a,b,c) contains cubic products of $a_1$, $b_j$, and $c_k$, but to the order of approximation implemented here this contribution can be expressed as a
sum of the eight quantities $a_1^3$, $a_1^2b_2$, $a_1b_2^2$, $a_1^2b_3$, $a_1^2c_3$, $b_2^3$, $a_1b_2b_3$, and $a_1b_2c_3$ times coefficients arising from the matrix
elements of their respective operators, multiplied by a global factor containing the integrals over the loops; in Table \ref{t:x3} we have listed these coefficients.
Finally, the contribution of Fig.~\ref{fig:eins}(d) can be expressed as a sum of the three parameters $a_1$, $b_2$, and $b_3$ times coefficients from the matrix elements
of the corresponding operators, also multiplied by a global factor containing the integrals over the loops. For completeness these coefficients can be found in Table
\ref{t:x4}. However, a few clarifying notes are instructive here. In Tables \ref{t:x3} and \ref{t:x4}, the singlet, octet, and \textbf{27} contributions are explicitly
separated so that the interested reader can reproduce our results. Besides, the singlet and octet pieces have been subtracted from the entries corresponding to the
\textbf{27} piece so that it is a purely \textbf{27} contribution. In order to simplify our notation, a coefficient that multiplies the entries of each flavor
representation has been factored out.

\begingroup
\begin{table}
\caption{\label{t:x2}Coefficients for the axial vector couplings of the baryons: tree level values.}
\begin{center}
\begin{tabular}{lccc}
\hline\hline
$B_jB_i$ & $a_1$ & $b_2$ & $b_3$ \\ \hline
$pn$ & 5/6 & 1/6 & 5/18 \\
$\Lambda \Sigma^\pm$ & $1/\sqrt{6}$ & 0 & $\sqrt{6}/18$ \\
$\Xi^0\Xi^-$ & 1/6 & $-1/6$ & 1/18 \\
$p\Lambda$ & $-\sqrt{3/8}$ & $-\sqrt{6}/12$ & $-\sqrt{6}/12$ \\
$n\Sigma^-$ & 1/6 & $-1/6$ & 1/18 \\
$\Lambda \Xi^-$ & $\sqrt{6}/12$ & $\sqrt{6}/12$ & $\sqrt{6}/36$ \\
$\Sigma^0\Xi^-$ & $5/\sqrt{72}$ & $\sqrt{2}/12$ & $5\sqrt{2}/36$ \\
$\Sigma^+\Xi^0$ & 5/6 & 1/6 & 5/18 \\
\hline\hline
\end{tabular}
\end{center}
\end{table}
\endgroup

\begingroup
\squeezetable
\begin{table}
\caption{\label{t:x3}Coefficients for the axial vector couplings of the baryons, Figs.~1(a,b,c).}
\begin{center}
\begin{tabular}{lcccccccccccccccccc}
\hline\hline
& \multicolumn{9}{c}{Singlet} \\ \hline
$B_jB_i$ & $C_{\mathbf{1}}^{B_jB_i}$ & $a_1^3$ & $a_1^2b_2$ & $a_1b_2^2$ & $a_1^2b_3$ & $a_1^2c_3$ &
$b_2^3$ & $a_1b_2b_3$ & $a_1b_2c_3$ \\ \hline
$pn$ &
1/432 & 345 & 63 & 171 & $-31$ & $-396$ & 21 & 338 & $-444$ \\
$\Lambda \Sigma^\pm$ &
$\sqrt{6}/432$ & $69$ & $-48$ & $15$ & $37$ & $-72$ & 0 & 40 & $-108$ \\
$\Xi^0\Xi^-$ &
$1/432$ & $69$ & $-351$ & $-81$ & $253$ & $-36$ & $-21$ & $-98$ & $-204$ \\
$p\Lambda$ &
$\sqrt{6}/288$ & $-69$ & $-53$ & $-47$ & $35$ & $84$ & $-7$ & $-86$ & 76 \\
$n\Sigma^-$ &
$1/432$ & 69$$ & $-351$ & $-81$ & $253$ & $-36$ & $-21$ & $-98$ & $-204$ \\
$\Lambda \Xi^-$ &
$\sqrt{6}/864$ & $69$ & $255$ & $111$ & $-179$ & $-108$ & 21 & 178 & $-12$ \\
$\Sigma^0\Xi^-$ &
$\sqrt{2}/864$ & $345$ & $63$ & $171$ & $-31$ & $-396$ & 21 & 338 & $-444$ \\
$\Sigma^+\Xi^0$ &
$1/432$ & $345$ & $63$ & $171$ & $-31$ & $-396$ & 21 & 338 & $-444$ \\
& \multicolumn{9}{c}{Octet} \\ \hline
$B_jB_i$ & $C_{\mathbf{8}}^{B_jB_i}$ & $a_1^3$ & $a_1^2b_2$ & $a_1b_2^2$ & $a_1^2b_3$ & $a_1^2c_3$ &
$b_2^3$ & $a_1b_2b_3$ & $a_1b_2c_3$ \\ \hline
$pn$ &
$\sqrt{3}/2592$ & 165 & $-381$ & $-33$ & $-419$ & $-564$ & $-15$ & 218 & $-708$ \\
$\Lambda \Sigma^\pm$ &
$\sqrt{2}/864$ & $33$ & $-96$ & $-9$ & $-71$ & $36$ & 0 & $-24$ & $-60$ \\
$\Xi^0\Xi^-$ &
$\sqrt{3}/2592$ & $33$ & $141$ & $147$ & $-407$ & $180$ & 15 & 86 & 12 \\
$p\Lambda$ &
$\sqrt{2}/3456$ & $99$ & $195$ & $105$ & $699$ & $540$ & 33 & $-118$ & 372 \\
$n\Sigma^-$ &
$\sqrt{3}/5184$ & $-33$ & $-141$ & $-147$ & $407$ & $-180$ & $-15$ & $-86$ & $-12$ \\
$\Lambda \Xi^-$ &
$\sqrt{2}/1152$ & $-11$ & $-241$ & $-97$ & $-147$ & $44$ & $-11$ & $-126$ & $-52$ \\
$\Sigma^0\Xi^-$ &
$\sqrt{6}/10368$ & $-165$ & $381$ & $33$ & $419$ & $564$ & 15 & $-218$ & 708 \\
$\Sigma^+\Xi^0$ &
$\sqrt{3}/5184$ & $-165$ & $381$ & $33$ & $419$ & $564$ & $15$ & $-218$ & 708 \\
& \multicolumn{9}{c}{27} \\ \hline
$B_jB_i$ & $C_{\mathbf{27}}^{B_jB_i}$ & $a_1^3$ & $a_1^2b_2$ & $a_1b_2^2$ & $a_1^2b_3$ & $a_1^2c_3$ &
$b_2^3$ & $a_1b_2b_3$ & $a_1b_2c_3$ \\ \hline
$pn$ &
1/5760 & 45 & 267 & 231 & $-107$ & $-92$ & 25 & 314 & $-204$ \\
$\Lambda \Sigma^\pm$ &
$\sqrt{6}/17280$ & $27$ & $-144$ & $-111$ & $-69$ & $264$ & 0 & $-296$ & 300 \\
$\Xi^0\Xi^-$ &
$1/5760$ & $9$ & $213$ & $-69$ & $609$ & $-340$ & $-25$ & 118 & 36 \\
$p\Lambda$ &
$\sqrt{6}/11520$ & $81$ & $225$ & $195$ & $-39$ & $-60$ & 27 & 238 & $-132$ \\
$n\Sigma^-$ &
$1/5760$ & $-27$ & $-159$ & $-33$ & $13$ & $-140$ & $-5$ & 46 & $-228$ \\
$\Lambda \Xi^-$ &
$\sqrt{6}/11520$ & $-27$ & $63$ & $-129$ & $381$ & $-132$ & $-27$ & $-62$ & 156 \\
$\Sigma^0\Xi^-$ &
$\sqrt{2}/11520$ & $-135$ & $-321$ & $27$ & $-719$ & $236$ & 5 & $-62$ & $-228$ \\
$\Sigma^+\Xi^0$ &
$1/5760$ & $-135$ & $-321$ & $27$ & $-719$ & $236$ & 5 & $-62$ & $-228$ \\
\hline\hline
\end{tabular}
\end{center}
\end{table}
\endgroup

\begingroup
\begin{table}
\caption{\label{t:x4}Coefficients for the axial vector couplings of the baryons. Fig.~1(d)}
\begin{center}
\begin{tabular}{lcccccccccccc}
\hline\hline
& \multicolumn{4}{c}{Singlet} & \multicolumn{4}{c}{Octet} & \multicolumn{4}{c}{27} \\ \hline
$B_jB_i$ &
$D_{\mathbf{1}}^{B_jB_i}$ & $a_1$ & $b_2$ & $b_3$ &
$D_{\mathbf{8}}^{B_jB_i}$ & $a_1$ & $b_2$ & $b_3$ &
$D_{\mathbf{27}}^{B_jB_i}$ & $a_1$ & $b_2$ & $b_3$ \\ \hline
$pn$ &
$-1/12$ & 15 & 3 & 5 & $-\sqrt{3}/72$ & 15 & 3 & 5 & 1/480 & 15 & 3 & 5 \\
$\Lambda \Sigma^\pm$ &
$-\sqrt{6}/12$ & 3 & 0 & 1 & $-\sqrt{2}/24$ & 3 & 0 & 1 & $\sqrt{6}/480$ & 3 & 0 & 1 \\
$\Xi^0\Xi^-$ &
$-1/12$ & 3 & $-3$ & 1 & $-\sqrt{3}/72$ & 3 & $-3$ & 1 & 1/480 & 3 & $-3$ & 1 \\
$p\Lambda$ &
$\sqrt{6}/8$ & 3 & 1 & 1 & $-\sqrt{2}/32$ & 3 & 1 & 1 & $3\sqrt{6}/320$ & 3 & 1 & 1 \\
$n\Sigma^-$ &
$-1/12$ & 3 & $-3$ & 1 & $\sqrt{3}/144$ & 3 & $-3$ & 1 & $-1/160$ & 3 & $-3$ & 1 \\
$\Lambda \Xi^-$ &
$-\sqrt{6}/24$ & 3 & 3 & 1 & $\sqrt{2}/96$ & 3 & 3 & 1 & $-\sqrt{6}/320$ & 3 & 3 & 1 \\
$\Sigma^0\Xi^-$ &
$-\sqrt{2}/24$ & 15 & 3 & 5 & $\sqrt{6}/288$ & 15 & 3 & 5 & $-\sqrt{2}/320$ & 15 & 3 & 5 \\
$\Sigma^+\Xi^0$ &
$-1/12$ & 15 & 3 & 5 & $\sqrt{3}/144$ & 15 & 3 & 5 & $-1/160$ & 15 & 3 & 5 \\
\hline\hline
\end{tabular}
\end{center}
\end{table}
\endgroup

Accordingly, for the process $n\to p e^- \overline{\nu}_e$ for instance, $g_A^{pn}$ can be constructed by reading off the appropriate
coefficients from Tables \ref{t:x2}-\ref{t:x4}, namely,
\begin{eqnarray}
g_A^{pn} & = & \alpha_{pn} \nonumber \\
&  & \mbox{} + C_\mathbf{1}^{pn}\left( 345a_1^3 + 63a_1^2b_2 + 171 a_1b_2^2 - 31a_1^2b_3 - 396a_1^2c_3+21b_2^3+338a_1b_2b_3-444a_1b_2c_3\right)F_\mathbf{1}^{(1)} \nonumber \\
&  & \mbox{} + C_\mathbf{8}^{pn}\left(165a_1^3 - 381a_1^2b_2 - 33a_1b_2^2 - 419a_1^2b_3 -564a_1^2c_3-15b_2^3+218a_1b_2b_3-708a_1b_2c_3\right)F_\mathbf{8}^{(1)} \nonumber \\
&  & \mbox{} + C_\mathbf{27}^{pn} \left(45a_1^3 + 267a_1^2b_2 + 231a_1b_2^2 -107a_1^2b_3 - 92a_1^2c_3+25b_2^3+314a_1b_2b_3-204a_1b_2c_3\right)F_\mathbf{27}^{(1)} \nonumber \\
&  & \mbox{} + D_\mathbf{1}^{pn} \left(15a_1 + 3b_2 + 5b_3\right)I_\mathbf{1} +
D_\mathbf{8}^{pn} \left(15a_1 + 3b_2 + 5b_3\right)I_\mathbf{8} + D_\mathbf{27}^{pn} \left(15a_1 + 3b_2 + 5b_3\right)I_\mathbf{27}, \nonumber \\
\end{eqnarray}
where the tree-level value reads
\begin{equation}
\alpha_{pn} = \frac56 a_1 + \frac16 b_2 + \frac{5}{18} b_3.
\end{equation}
Note that the coefficients $C_\mathbf{1}^{pn}=1/432$, $C_\mathbf{8}^{pn}=\sqrt{3}/2592$, $C_\mathbf{27}^{pn}=1/5760$, $D_\mathbf{1}^{pn}=-1/12$,
$D_\mathbf{8}^{pn}=-\sqrt{3}/72$, and $D_\mathbf{27}^{pn}= 1/480$ are the common factors that multiply each entry referred to above.
Analogous expressions can be obtained for the axial vector couplings of the remaining processes.

\section{Chiral coefficients\label{appB}}

In this Appendix, for completeness, the explicit formulas for the chiral coefficients introduced in Eq.~(\ref{eq:axren}) are given.

The lowest order coefficients $\alpha_{B_jB_i}$ are
\begin{eqnarray}
\begin{array}{ll}
\displaystyle
\alpha_{pn} = D + F, \qquad &
\displaystyle
\alpha_{\Lambda \Sigma^\pm} = \frac{2}{\sqrt 6} D, \\ [3mm]
\displaystyle
\alpha_{p \Lambda} = - \frac{1}{\sqrt 6} (D + 3 F), \qquad &
\displaystyle
\alpha_{n \Sigma^-} = D - F, \\ [3mm]
\displaystyle
\alpha_{\Lambda \Xi^-} = - \frac{1}{\sqrt 6} (D - 3 F),
\qquad &
\displaystyle
\alpha_{\Xi^0 \Xi^-} = D - F, \\ [3mm]
\displaystyle
\alpha_{\Sigma^0 \Xi^-} = \frac{1}{\sqrt 2} (D + F) = \frac{1}{\sqrt 2} \alpha_{\Sigma^+ \Xi^0}. \qquad &
\end{array} \nonumber
\end{eqnarray}

The coefficients $\bar{\lambda}_{B_i}^\Pi$ arising from the one-loop correction due to wavefunction renormalization,
Figs.~\ref{fig:eins}(b,c), are for the octet baryons
\begin{eqnarray}
\begin{array}{ll}
\displaystyle
\bar{\lambda}_N^\pi = \frac94 {(F + D)}^2 + 2 \mathcal{C}^2, \qquad &
\displaystyle
\bar{\lambda}_\Sigma^\pi = 6 F^2 + D^2 + \frac13 \mathcal{C}^2, \\ [3mm]
\displaystyle
\bar{\lambda}_N^K = \frac12 (9 F^2 - 6 F D + 5 D^2 + \mathcal{C}^2), \qquad &
\displaystyle
\bar{\lambda}_\Sigma^K = 3 (F^2 + D^2) + \frac53 \mathcal{C}^2, \\ [3mm]
\displaystyle
\bar{\lambda}_N^\eta = \frac14 {(3 F - D)}^2, \qquad &
\displaystyle
\bar{\lambda}_\Sigma^\eta = D^2 + \frac12 \mathcal{C}^2, \\ [3mm]
\displaystyle
\bar{\lambda}_\Xi^\pi = \frac94 {(F - D)}^2 + \frac12 \mathcal{C}^2, \qquad &
\displaystyle
\bar{\lambda}_\Lambda^\pi = 3 D^2 + \frac32 \mathcal{C}^2, \\ [3mm]
\displaystyle
\bar{\lambda}_\Xi^K = \frac12 (9 F^2 + 6 F D + 5 D^2 + 3 \mathcal{C}^2), \qquad &
\displaystyle
\bar{\lambda}_\Lambda^K = 9 F^2 + D^2 + \mathcal{C}^2, \\ [3mm]
\displaystyle
\bar{\lambda}_\Xi^\eta =  \frac14 {(3 F + D)}^2 + \frac12 \mathcal{C}^2, \qquad &
\displaystyle
\bar{\lambda}_\Lambda^\eta = D^2.\label{eq:lreno}
\end{array} \nonumber
\end{eqnarray}
and for the decuplet baryons
\begin{eqnarray}
\begin{array}{ll}
\displaystyle
\bar{\lambda}_\Delta^\pi = \frac{25}{36} \mathcal{H}^2  + \frac12 \mathcal{C}^2, \qquad &
\displaystyle
\bar{\lambda}_{\Xi^*}^\pi = \frac{5}{36} \mathcal{H}^2 + \frac14 \mathcal{C}^2, \\ [3mm]
\displaystyle
\bar{\lambda}_\Delta^K = \frac{5}{18} \mathcal{H}^2 + \frac12 \mathcal{C}^2, \qquad &
\displaystyle
\bar{\lambda}_{\Xi^*}^K = \frac{5}{6} \mathcal{H}^2 + \frac12 \mathcal{C}^2, \\ [3mm]
\displaystyle
\bar{\lambda}_\Delta^\eta = \frac{5}{36} \mathcal{H}^2, \qquad &
\displaystyle
\bar{\lambda}_{\Xi^*}^\eta =  \frac{5}{36} \mathcal{H}^2 + \frac14 \mathcal{C}^2, \\ [3mm]
\displaystyle
\bar{\lambda}_{\Sigma^*}^\pi = \frac{10}{27} \mathcal{H}^2 + \frac{5}{12} \mathcal{C}^2, \qquad &
\displaystyle
\bar{\lambda}_{\Omega^-}^\pi = \frac{10}{27} \mathcal{H}^2, \\ [3mm]
\displaystyle
\bar{\lambda}_{\Sigma^*}^K =  \frac{20}{27} \mathcal{H}^2 + \frac13 \mathcal{C}^2, \qquad &
\displaystyle
\bar{\lambda}_{\Omega^-}^K = \frac59 \mathcal{H}^2 + \mathcal{C}^2, \\ [3mm]
\displaystyle
\bar{\lambda}_{\Sigma^*}^\eta = \frac14 \mathcal{C}^2, \qquad &
\displaystyle
\bar{\lambda}_{\Omega^-}^\eta =  \frac59 \mathcal{H}^2 + \mathcal{C}^2.
\end{array} \nonumber
\end{eqnarray}

The coefficients $\bar{\lambda}_{B_jB_i}^\Pi$ are thus written as
\begin{eqnarray}
\begin{array}{ll}
\displaystyle
\bar{\lambda}_{pn}^\pi = \frac94 {(F + D)}^2 + 2 \mathcal{C}^2, \quad &
\displaystyle
\bar{\lambda}_{\Lambda \Sigma^\pm}^\pi = 3 F^2 + 2 D^2 + \frac{11}{12} \mathcal{C}^2, \\ [3mm]
\displaystyle
\bar{\lambda}_{pn}^K = \frac12 (9 F^2 - 6 F D + 5 D^2 + \mathcal{C}^2), \quad &
\displaystyle
\bar{\lambda}_{\Lambda \Sigma^\pm}^K = 6 F^2 + 2 D^2 + \frac43 \mathcal{C}^2, \\ [3mm]
\displaystyle
\bar{\lambda}_{pn}^\eta = \frac14 {(3 F - D)}^2, \quad &
\displaystyle
\bar{\lambda}_{\Lambda \Sigma^\pm}^\eta = D^2 + \frac14 \mathcal{C}^2, \\ [3mm]
\displaystyle
\bar{\lambda}_{p \Lambda}^\pi = \frac38 (3 F^2 + 6 FD + 7 D^2) + \frac74 \mathcal{C}^2, \quad &
\displaystyle
\bar{\lambda}_{n \Sigma^-}^\pi = \frac18 (33 F^2 + 18 F D + 13 D^2) + \frac76 \mathcal{C}^2, \\ [3mm]
\displaystyle
\bar{\lambda}_{p \Lambda}^K = \frac14 (27 F^2 - 6 F D + 7 D^2) + \frac34 \mathcal{C}^2, \quad &
\displaystyle
\bar{\lambda}_{n \Sigma^-}^K = \frac14 (15 F^2 - 6 F D + 11 D^2) + \frac{13}{12} \mathcal{C}^2, \\ [3mm]
\displaystyle
\bar{\lambda}_{p \Lambda}^\eta = \frac18 (9 F^2 - 6 F D + 5 D^2), \quad &
\displaystyle
\bar{\lambda}_{n \Sigma^-}^\eta = \frac18 (9 F^2 - 6 F D + 5 D^2) + \frac14 \mathcal{C}^2, \\ [3mm]
\displaystyle
\bar{\lambda}_{\Lambda \Xi^-}^\pi = \frac38 (3 F^2 - 6 F D + 7 D^2) + \mathcal{C}^2, \quad &
\displaystyle
\bar{\lambda}_{\Sigma^0 \Xi^-}^\pi = \frac18 (33 F^2 - 18 F D + 13 D^2) + \frac{5}{12} \mathcal{C}^2, \\ [3mm]
\displaystyle
\bar{\lambda}_{\Lambda \Xi^-}^K = \frac14 (27 F^2 + 6 F D + 7 D^2) + \frac54 \mathcal{C}^2, \quad &
\displaystyle
\bar{\lambda}_{\Sigma^0 \Xi^-}^K = \frac14 (15 F^2 + 6 F D + 11 D^2) + \frac{19}{12} \mathcal{C}^2, \\ [3mm]
\displaystyle
\bar{\lambda}_{\Lambda \Xi^-}^\eta = \frac18 (9 F^2 + 6 F D + 5 D^2) + \frac14 \mathcal{C}^2, \quad &
\displaystyle
\bar{\lambda}_{\Sigma^0 \Xi^-}^\eta = \frac18 (9 F^2 + 6 F D + 5 D^2) + \frac12 \mathcal{C}^2, \\ [3mm]
\displaystyle
\bar{\lambda}_{\Xi^0 \Xi^-}^\pi = \frac94 {(F - D)}^2 + \frac12 \mathcal{C}^2, \quad &
\displaystyle
\bar{\lambda}_{\Sigma^+ \Xi^0}^\pi = \bar{\lambda}_{\Lambda \Xi^-}^\pi, \\ [3mm]
\displaystyle
\bar{\lambda}_{\Xi^0 \Xi^-}^K = \frac12 (9 F^2 + 6 F D + 5 D^2) + \frac32 \mathcal{C}^2, \quad &
\displaystyle
\bar{\lambda}_{\Sigma^+ \Xi^0}^K = \bar{\lambda}_{\Lambda \Xi^-}^K, \\ [3mm]
\displaystyle
\bar{\lambda}_{\Xi^0 \Xi^-}^\eta = \frac14 {(3 F + D)}^2 + \frac12 \mathcal{C}^2, \quad &
\displaystyle
\bar{\lambda}_{\Sigma^+ \Xi^0}^\eta = \bar{\lambda}_{\Lambda \Xi^-}^\eta.
\end{array} \nonumber
\end{eqnarray}

The coefficients $\bar{\beta}_{B_jB_i}^\Pi$ evaluated from the graph in Fig.~\ref{fig:eins}(a) are
\begin{eqnarray}
\bar{\beta}_{pn}^\pi & = & \frac14 {(F + D)}^3 + \frac{16}{9} (F + D) \mathcal{C}^2 - \frac{50}{81} \mathcal{H}
\mathcal{C}^2, \nonumber \\
\bar{\beta}_{pn}^K & = & \frac13 (-3 F^3 + 3 F^2 D - F D^2 + D^3) + \frac29 (F + 3 D) \mathcal{C}^2 - \frac{10}{81} \mathcal{H} \mathcal{C}^2, \nonumber \\
\bar{\beta}_{pn}^{\eta} & = & - \frac{1}{12} (F + D){(3 F - D)}^2, \nonumber \\
\bar{\beta}_{\Lambda \Sigma^\pm}^\pi & = & \frac{2}{3 \sqrt 6} D (6 F^2 - D^2) + \frac{2}{3 \sqrt 6} (2 F + \frac13 D) \mathcal{C}^2 - \frac{10}{27 \sqrt 6} \mathcal{H}
\mathcal{C}^2, \nonumber \\
\bar{\beta}_{\Lambda \Sigma^\pm}^{K} & = & - \frac{1}{\sqrt 6} D (F^2 - D^2) + \frac{8}{3 \sqrt 6} (F + \frac23 D) \mathcal{C}^2 - \frac{5}{27 \sqrt6} \mathcal{H}
\mathcal{C}^2, \nonumber \\
\bar{\beta}_{\Lambda \Sigma^\pm}^{\eta} & = & \frac{2}{3 \sqrt 6} D (D^2 + \mathcal{C}^2), \nonumber \\
\bar{\beta}_{p \Lambda}^\pi & = & \frac{3}{2 \sqrt 6} D (F^2 - D^2) - \frac{1}{3 \sqrt 6} (11 D + 3 F) \mathcal{C}^2 + \frac{10}{9 \sqrt 6} \mathcal{H} \mathcal{C}^2,
\nonumber \\
\bar{\beta}_{p \Lambda}^K & = & \frac{1}{6 \sqrt 6} (27 F^3 - 9 F^2 D - 15 F D^2 + 5 D^3) - \frac{1}{\sqrt 6} (F + D) \mathcal{C}^2 + \frac{5}{9 \sqrt 6} \mathcal{H}
\mathcal{C}^2, \nonumber \\
\bar{\beta}_{p \Lambda}^{\eta} & = & - \frac{1}{6 \sqrt 6} D (9 F^2 - D^2), \nonumber \\
\bar{\beta}_{n \Sigma^-}^\pi & = & \frac16 (6 F^3 + 3 F^2 D - 2 F D^2 + D^3) + \frac29 (5 F + D) \mathcal{C}^2 + \frac{10}{81} \mathcal{H} \mathcal{C}^2, \nonumber \\
\bar{\beta}_{n \Sigma^-}^K & = & \frac16 (3 F^3 + 3 F^2 D + F D^2 + D^3) + \frac19 (5 F + D) \mathcal{C}^2 + \frac{5}{81} \mathcal{H} \mathcal{C}^2, \nonumber \\
\bar{\beta}_{n \Sigma^-}^{\eta} & = & \frac16 D (3 F^2 - 4 F D + D^2) + \frac19 (3 F - D) \mathcal{C}^2, \nonumber \\
\bar{\beta}_{\Lambda \Xi^-}^\pi & = & \frac{3}{2 \sqrt 6} D (F^2 - D^2) - \frac{1}{3 \sqrt 6} (3 F - D) \mathcal{C}^2 - \frac{5}{9 \sqrt 6} \mathcal{H} \mathcal{C}^2,
\nonumber \\
\bar{\beta}_{\Lambda \Xi^-}^K & = & \frac{1}{6 \sqrt 6} (- 27 F^3 - 9 F^2 D + 15 F D^2 + 5 D^3) - \frac{1}{\sqrt 6} (F - D) \mathcal{C}^2 - \frac{5}{9 \sqrt 6} \mathcal{H}
\mathcal{C}^2, \nonumber \\
\bar{\beta}_{\Lambda \Xi^-}^{\eta} & = & - \frac{1}{6 \sqrt 6} D (9 F^2 - D^2) + \frac{2}{3 \sqrt 6} D \mathcal{C}^2, \nonumber \\
\bar{\beta}_{\Sigma^0 \Xi^-}^\pi & = & \frac{1}{6 \sqrt 2} (-6F^3 + 3 F^2 D + 2 F D^2 + D^3) + \frac{2}{9 \sqrt 2} (F + 2 D) \mathcal{C}^2 - \frac{10}{81 \sqrt 2}
\mathcal{H} \mathcal{C}^2 \nonumber \\
\bar{\beta}_{\Sigma^0 \Xi^-}^K & = & \frac{1}{6 \sqrt 2} (- 3 F^3 + 3 F^2 D - F D^2 + D^3) + \frac{1}{9 \sqrt 2} (13 F + 15 D) \mathcal{C}^2 - \frac{35}{81 \sqrt 2}
\mathcal{H} \mathcal{C}^2 \nonumber \\
\bar{\beta}_{\Sigma^0 \Xi^-}^{\eta} & = & \frac{1}{6 \sqrt 2} D(3 F^2 + 4 F D + D^2) + \frac{1}{3 \sqrt 2} (F + D) \mathcal{C}^2 - \frac{5}{27 \sqrt 2} \mathcal{H}
\mathcal{C}^2 \nonumber \\
\bar{\beta}_{\Xi^0 \Xi^-}^\pi & = & - \frac14 {(F - D)}^3 + \frac29 (F - D) \mathcal{C}^2 - \frac{5}{162} \mathcal{H} \mathcal{C}^2, \nonumber \\
\bar{\beta}_{\Xi^0 \Xi^-}^K & = & \frac13 (3 F^3 + 3 F^2 D + F D^2 + D^3) + \frac29 (5 F + D) \mathcal{C}^2 + \frac{10}{81} \mathcal{H} \mathcal{C}^2, \nonumber \\
\bar{\beta}_{\Xi^0 \Xi^-}^{\eta} & = & \frac{1}{12} (F - D){(3 F + D)}^2 + \frac29 (3 F + D) \mathcal{C}^2  + \frac{5}{54} \mathcal{H} \mathcal{C}^2. \nonumber
\end{eqnarray}
and, due to isospin symmetry, one also has
\begin{equation*}
\bar{\beta}_{\Sigma^+ \Xi^0}^\Pi = \sqrt{2} \, \bar{\beta}_{\Sigma^0 \Xi^-}^\Pi. \qquad (\Pi=\pi,K,\eta)
\end{equation*}

Now, the coefficients $\gamma_{B_jB_i}^\Pi$ from Fig.~\ref{fig:eins}(d) are
\begin{eqnarray}
\begin{array}{ll}
\displaystyle
\gamma_{pn}^\pi = - F - D, \qquad &
\displaystyle
\gamma_{\Lambda \Sigma^\pm}^\pi = -\frac{2}{\sqrt 6} D, \\ [3mm]
\displaystyle
\gamma_{pn}^{K} = - \frac12 (F + D), \qquad &
\displaystyle
\gamma_{\Lambda\Sigma^\pm}^K = - \frac{1}{\sqrt 6} D, \\ [3mm]
\displaystyle
\gamma_{pn}^\eta = 0, \qquad &
\displaystyle
\gamma_{\Lambda \Sigma^\pm}^\eta = 0, \\ [3mm]
\displaystyle
\gamma_{p \Lambda}^\pi = \frac{3}{8 \sqrt 6} ( 3F + D), \qquad &
\displaystyle
\gamma_{n \Sigma^-}^\pi = \frac38 (F - D), \\ [3mm]
\displaystyle
\gamma_{p \Lambda}^K =  \frac{3}{4 \sqrt 6} ( 3F + D), \qquad &
\displaystyle
\gamma_{n \Sigma^-}^K = \frac34 (F - D), \\ [3mm]
\displaystyle
\gamma_{p \Lambda}^{\eta} = \frac{3}{8 \sqrt 6} (3F + D), \qquad &
\displaystyle
\gamma_{n \Sigma^-}^{\eta} = \frac38 (F - D), \\ [3mm]
\displaystyle
\gamma_{\Lambda \Xi^-}^\pi = - \frac{3}{8 \sqrt 6} (3F-D), \qquad &
\displaystyle
\gamma_{\Sigma^0 \Xi^-}^\pi = - \frac{3}{8 \sqrt 2} (F + D), \\ [3mm]
\displaystyle
\gamma_{\Lambda \Xi^-}^K = - \frac{3}{4 \sqrt 6} (3F - D), \qquad &
\displaystyle
\gamma_{\Sigma^0 \Xi^-}^K =   - \frac{3}{4 \sqrt 2} (F + D), \\ [3mm]
\displaystyle
\gamma_{\Lambda \Xi^-}^{\eta} = - \frac{3}{8 \sqrt 6} (3F - D), \qquad &
\displaystyle
\gamma_{\Sigma^0 \Xi^-}^{\eta} = - \frac{3}{8 \sqrt 2} (F + D), \\ [3mm]
\displaystyle
\gamma_{\Xi^0 \Xi^-}^\pi = F - D, \qquad &
\displaystyle
\gamma_{\Sigma^+ \Xi^0}^\pi = - \frac38 (F + D), \\ [3mm]
\displaystyle
\gamma_{\Xi^0 \Xi^-}^K = \frac12 (F - D), \qquad &
\displaystyle
\gamma_{\Sigma^+ \Xi^0}^K =  - \frac34 (F + D), \\ [3mm]
\displaystyle
\gamma_{\Xi^0 \Xi^-}^{\eta} = 0, \qquad &
\displaystyle
\gamma_{\Sigma^+ \Xi^0}^{\eta} =  - \frac38 (F + D).
\end{array} \nonumber
\end{eqnarray}

The chiral coefficients listed above include contributions from intermediate octet and decuplet baryons. The corresponding distinction between primed and unprimed
coefficients as defined in Eq.~(\ref{eq:axren}) is straightforward.

Finally, the coefficients $\zeta_{B_jB_i}^{\eta^\prime}$ from Fig.~\ref{fig:eins}(a,b,c) read

\begin{eqnarray}
&  & \zeta_{pn}^{\eta^\prime} = \frac19 (F+D)(3F-D)^2, \nonumber \\
&  & \zeta_{\Lambda \Sigma^\pm}^{\eta^\prime} = \frac19 \sqrt{\frac{2}{3}} D(3F-D)^2, \nonumber \\
&  & \zeta_{\Xi^0\Xi^-}^{\eta^\prime} = \frac19 (D-F)(3F-D)^2, \nonumber \\
&  & \zeta_{p\Lambda}^{\eta^\prime} = -\frac{1}{9\sqrt 6} (3F+D)(3F-D)^2, \nonumber \\
&  & \zeta_{n\Sigma^-}^{\eta^\prime} = \frac19 (D-F)(3F-D)^2, \nonumber \\
&  & \zeta_{\Lambda \Xi^-}^{\eta^\prime} =  \frac{1}{9\sqrt 6}(3F-D)^3, \nonumber \\
&  & \zeta_{\Sigma^0\Xi^-}^{\eta^\prime} = \frac{1}{9\sqrt{2}} (F+D)(3F-D)^2, \nonumber \\
&  & \zeta_{\Sigma^+\Xi^0}^{\eta^\prime} = \frac19 (F+D)(3F-D)^2. \nonumber
\end{eqnarray}

\end{document}